\documentclass[letterpaper]{JHEP3}

\usepackage{graphicx,epsf,amssymb,amsbsy,amsfonts,amssymb,amsmath,amsthm}

\newcommand{\da}{{\dot{a}}}

\def\nn{\nonumber}

\def\s#1{{      
        \setbox\charbox=\hbox{$#1$}
        \setbox\slabox=\hbox{$/$}
        \dimen\charbox=\ht\slabox
        \advance\dimen\charbox by -\dp\slabox
        \advance\dimen\charbox by -\ht\charbox
        \advance\dimen\charbox by \dp\charbox
        \divide\dimen\charbox by 2
        \raise-\dimen\charbox\hbox to \wd\charbox{\hss/\hss}
        \llap{$#1$} }}

\title{Non-Chiral $\mathcal{S}$-Matrix of $\mathcal{N}=4$ Super Yang-Mills}
\author{Yu-tin Huang\footnote{Email: yhuang@physics.ucla.edu}
\\ \\
\\
\it  Department of Physics and Astronomy,\\ University of California at Los Angeles,\\
CA 90095-1547, USA\;\\
\;
\\
\it  Kavli Institute for Theoretical Physics,\\  University of California at Santa Barbara,\\
CA 93106-4030, USA\;}
\abstract{We discuss the construction of non-chiral $S$ matrix of four-dimensional $\mathcal{N}=4$ super Yang-Mills using a non-chiral superspace. This construction utilizes the non-chiral representation of dual superconformal symmetry, which is the natural representation from the point of view of the six-dimensional parent theory. The superspace in discussion is projective superspace constructed by Hatsuda and Siegel, and is based on a half coset U(2,2$|$4)/U(1,1$|$2)$^2_+$. We obtain the non-chiral representation of the five-point and general $n$-point MHV-$\overline{\rm MHV}$ amplitude. The non-chiral formulation can be straightforwardly lifted to six dimensions, which is equivalent to massive amplitudes in four dimensions.  }
\preprint{ UCLA-TEP-11-106\; NSF-KITP-11-048}

\begin{document}

\section{Introduction}
In recent years, newly discovered symmetries and dualities have played an important role in the construction of S-matrix. The most prominent are the dual superconformal symmetry~\cite{DualConformal, DualConformal2} of planar $\mathcal{N}=4$ super Yang-Mills, and the color-kinematics duality for general Yang-Mills amplitudes~\cite{BCJ}. The former is a superconformal symmetry defined in an auxiliary dual space and has been established as a symmetry of the planar theory both at strong~\cite{Strong,FermiT1,FermiT2} and weak~\cite{Weak,anomaly,Elvang:2009ya} coupling. The generators of this new symmetry, when written in terms of the original on-shell variables, becomes an Yangian algebra~\cite{Drummond:2009fd}, and has played a central role in determining the S-matrix elements to all order in the number of external legs at tree-~\cite{Drummond:2008cr}, and at loop-level~\cite{Bern:2006ew, Korchemsky:2010ut, ArkaniHamed:2010kv}. The later, proposed by Bern, Carrasco and Johansson (BCJ), is the conjecture of the existence of a representation for (super)Yang-Mills S-matrix elements, where both the color and kinematic structure satisfies the same Jacobi relations. It was proven~\cite{Bern:2010yg} that once such a representation is found, one can directly obtain the (super)gravity S-matrix elements directly from (super)Yang-Mills ones. Such a relationship has been demonstrated up to three loops~\cite{BCJLoop}.

In this paper we will focus on a non-chiral version of dual superconformal (DSC) symmetry. Dual conformal symmetry has recently been established for maximal super Yang-Mills at tree-level in ten~\cite{CaronHuot:2010rj} and six dimensions~\cite{Bern:2010qa,Dennen:2010dh}, while the analysis in six-dimensions has also been extended to loop integrands due to the availability of supersymmetrized~\cite{DHS} spinor- helicity formalism~\cite{CheungOConnell,Boels:2009bv}. The lower dimensional dual symmetry can then be viewed as inherited from higher dimensions. In four dimensions, on-shell superconformal symmetry is most conveniently analysed using twistorial constructions. Indeed the current studies of dual superconformal symmetry~\cite{ArkaniHamed:2009vw, Mason:2009qx} utilizes the notion of momentum twistors~\cite{Hodges:2009hk}, which are twistor variables that defines the dual momentum spacetime. Such a construction is inherently chiral, reflected in the fact that the dual space is a chiral superspace. However, from the higher dimensional point of view, the natural space to establish such a symmetry is a non-chiral superspace. This can been seen by looking at six-dimensions, where the on-shell superfield for maximal super Yang-Mills is a scalar superfield~\cite{DHS}(see also \cite{Boels:2009bv}), and the grassmann variables transform under the non-chiral  SU(2)$\times$SU(2) R-symmetry. Thus the dual conformal symmetry in higher dimensions implies that there exists an alternative representation of the four dimensional DSC symmetry, where the superspace is non-chiral. 

There are various reasons why non-chiral formulations are desirable. From the practical view point, once the non-chiral formulation of the massless amplitude is obtained, one can straightforwardly obtain the massive amplitudes. More precisely, at tree-level the six dimensional parent superamplitude and its four dimensional massless reduction(all external lines lies in a four dimensional subplane) takes the same form (see ~\cite{Bern:2010qa,Elvang} for details). This is due to the fact that the six dimensional superamplitude is non-chiral, and hence Levi-Cevita tensors, which will vanish in four dimensions and potentially invalidate the equivalence, will not appear. Thus, at tree level, obtaining the four dimensional non-chiral amplitude is equivalent to obtaining the six dimensional form. After uplifting to six dimensions, using a massive reduction, i.e. an arbitrary massless momenta in six dimensions is equivalent to massive momenta in four dimensions, one obtains the four dimensional massive amplitude~\cite{Bern:2010qa}. The fact that the non-chiral massless amplitude knows about the massive extension, is really the power of higher-dimensional dual conformal symmetry, which constrains how the extra dimensional kinematics, or masses, can be introduced. In fact it is well known that the massively regulated amplitudes of $\mathcal{N}=4$ SYM are dual conformal invariant if one extends the dual symmetry to five dimensions and interpret the mass as the momenta of the fifth dimension~\cite{Alday:2009zm}.

From the theoretical view point, non-chiral amplitudes are potentially relevant for alternative generalizations of the Wilson loop-amplitude duality~\cite{Strong} beyond the MHV sector. A recent proposal, the correlator-Wilson loop-amplitudes duality~\cite{Alday:2010zy,Next}, is formulated in a non-chiral fashion. When compared the dual object to the amplitude, one by hand set the non-chiral part to zero. From a pure structural standpoint, it would be more natural to compare non-chiral objects to non-chiral objects, and hence a reformulation of the amplitudes in some non-chiral superspace will be desirable. 

The purpose of this letter is to demonstrate that such a non-chiral formulation can be constructed most easily by utilizing the non-chiral version of the dual symmetry. The dual symmetry is defined in an on-shell non-chiral superspace which is a half-coset superspace U(2,2$|$4)/U(1,1$|$2)$^2_+$ constructed by Siegel and Hatsuda~\cite{Hatsuda:2002wf,Hatsuda:2007wr}. There are eight bosonic and eight fermionic coordinates in this superspace, ($x_{\dot\mu}\,^\nu,\theta_{m'}\,^\nu,\bar{\theta}_{\dot\mu}\,^{n},y_{m'}^n$). This differs from the usual chiral superspace in that  fermionic coordinates of both chirality exists and the R-symmetry is broken down to SU(2)$\times$SU(2). The additional $y$ coordinates is a grassmann even coordinate, and related to the R-symmetry generators. Extending the DSC generators to the on-shell space, one finds besides the usual superconformal generators, 16 level-1 Yangian generators 
\begin{equation}
(K_{\dot\mu}\,^{\nu}\rightarrow J^{(1)}_{\dot\mu}\,^{\nu}, \bar{S}_{\dot\mu}\,^{n}\rightarrow\,J^{(1)}_{\dot\mu}\,^{n},\, S_{m'}\,^{\nu}\rightarrow J^{(1)}_{m'}\,^{\nu},R_{m'}\,^{n}\rightarrow\,J^{(1)}_{m'}\,^{n})\,.
\end{equation}
The remaining 16 level-1 generators can be obtained by (anti)commuting the above with the superconformal generators, which serve as the level zero generators of the Yangian algebra.

Dual superconformal symmetry of the planar amplitude implies that it can be written as a function of these coordinates. For example the four and five-point amplitude is given by 
\begin{eqnarray}
\nonumber\mathcal{A}_4&=&\delta^4(x_1-x_{5})\delta^4(\theta_1-\theta_{5})\delta^4(\bar\theta_1-\bar\theta_{5})\frac{1}{x^2_{13}x^2_{24}}\\
\mathcal{A}_5&=&\frac{\delta^4(x_1-x_{6})\delta^4(\theta_1-\theta_{6})\delta^4(\bar\theta_1-\bar\theta_{6})}{x^2_{13}x^2_{24}x^2_{35}x^2_{41}x^2_{52}}\left[\frac{\Theta_{5,24}x_{51}x_{12}x_{23}x_{34}x_{41}x_{15}\Theta_{5,24}}{x^2_{52}x^2_{35}x^2_{24}}\right]
\end{eqnarray} 
where $x_{ij}\equiv x_i-x_j$ and $\Theta_{5,24}\equiv (\theta+\bar\theta)_2x_{24}x_{45}+(\theta+\bar\theta)_4x_{42}x_{25}+(\theta+\bar\theta)_5x^2_{24}$.\footnote{The function $\Theta_{i,jk}$ is simply a non-chiral version of the dual conformal covariant object $\Xi_{i,jk}=(\theta_jx_{jk}x_{ki}+\theta_kx_{kj}x_{ji}+\theta_ix^2_{jk})|i\rangle$ defined in \cite{DualConformal2}} Note that the five point amplitude contains a non-trivial dual conformal covariant function in the square brackets, and can serve as building blocks of higher point amplitudes via BCFW recursion relation~\cite{BCFW}. The above results is derived by performing half-Fourier transformation of the (anti)-chiral representation. Such transformation can be straightforwardly carried out for $\overline{\rm MHV}$(MHV) amplitudes for all-$n$:
\begin{eqnarray}
\nonumber\mathcal{A}_n(\overline{\rm MHV})&=&\frac{\delta^4(x_{1(n+1)})\delta^4(\theta_{1(n+1)})\delta^4(\bar\theta_{1(n+1)})}{\prod_{i=1}^n[ ii+1]}\left[ \frac{\left(\langle n|x_{n,n-1}x_{n-1,2}\theta_{2,n}+\langle n|x_{n,2}x_{2,n-1}\theta_{n-1,n}\right)^2}{(\langle n1\rangle\langle nn-1\rangle\langle n-2n-3\rangle)^2}\right.   \\
\nonumber&&\left.\times\prod_{r=1}^{n-5}\frac{\left(\langle\xi_r|x_{r+1,n-5+r}x_{n-5+r,n-1}\theta_{n-1,r+1}+\langle\xi_r|x_{r+1,n-1}x_{n-1,n-5+r}\theta_{n-5+r,r+1}\right)^2}{(\langle\xi_r(1+r)\rangle x^2_{n-1,r+1})^2} \right]\,.\end{eqnarray}
Detailed definition of the notations are given in section \ref{section4}. The spinor products in the above expression are to be removed by multiplying and dividing by its anti-holomorphic partner.

This paper is organized as follows: In the next section we will give a lightning review of projective superspace in the language of half-cosets. We will derive the coordinates which parameterize the space, and give the generators of superconformal group in terms of these coordinates. In section \ref{section3}, we define the relationship between the on-shell space and the dual space, and deduce how the dual generators act on the on-shell variables. We will see that among the 64 dual generators, 16 becomes trivial, 32 becomes part of the original superconformal generators, and the remaining 16 becomes the level-1 generators. In section \ref{section4} we first discuss the relationship between the four- and six-dimensional tree amplitude and the structure of the amplitudes in the dual space, which is naturally separated into a (super)momentum conservation piece and a dual conformal covariant piece which is similar to the ``R"-function in the chiral approach. We will explicitly construct the five point amplitude and six point $\overline{\rm MHV}$ amplitude via half-Fourier transform. The transformed result for general-$n$ is given. With the use of ambi-momentum twistor space, we will expose the cyclic symmetry of the amplitude, which is non-manifest in the dual space coordinates. In section \ref{section5} we give the conclude and a discussion of future directions. 
\section{Projective superspace and superconformal generators. }
Here we give a quick introduction to half-coset spaces and in particular the half-coset U(2,2$|$4)/ U(1,1$|$2)$^2_+$, more details can be found in~\cite{Hatsuda:2002wf,Hatsuda:2007wr}. Both coset and half-cosets can be understood in the following unified construction. One begins by choosing a particular U(1) generator of the group $G$ and divide all the generators into those with positive $G_+$, negative $G_-$ and vanishing eigenvalues $G_0$. The generators $G_0$, which includes the chosen U(1), forms the isotropy group for the usual coset description. For half-cosets, one enlarges the isotropy group to include $G_-$ and thus leaving behind $G_+$, which defines the parameters of the space. Half cosets are conveniently labelled as $G/G_0+$, where the ``+" indicates that one quotients everything except the $G_+$ generators.

We now consider the particular example U(2,2$|$4)/U(1,1$|$2)$^2_+$ in more detail. We introduce the U(2,2$|$4) homogenous coordinates:
\begin{equation}
\nonumber \mathcal{Z}_{\mathcal{A}}\,^{\mathcal{M}}\quad\begin{array}{ccc} \mathcal{M}=M,\; M'  \\ \mathcal{A}=A,\;A'\end{array}
 \end{equation}
where the $U(2,2|4)$ isometry group index $\mathcal{M}$ is split into half with $M=\left(\mu,m\right), M'=\left(\dot\mu, m'\right)$ to match with the structure of the two $U(1,1|2)$ isotropy group index $\mathcal{A}=\left(A,A'\right)$, with $A=\left(\alpha,a\right)$ and $A'=\left(\dot\alpha,a'\right)$. The superspace coordinates are acted on the right by the isometry group generators 
\begin{equation}
G_{\mathcal{N}}\,^{\mathcal{M}}=\mathcal{Z}_{\mathcal{A}}\,^{\mathcal{M}}\frac{\partial}{\partial \mathcal{Z}_{\mathcal{A}}\,^{\mathcal{N}} }\,.
\end{equation}
For convenience we explicitly write out the superconformal generators in this construction: 
\begin{eqnarray}
\nonumber&&\quad\quad\quad\quad\quad\dot{\nu}\quad\quad\;\;\quad n'\quad\quad\;\; n\quad\quad\quad\nu\\
 G_{\mathcal{N}}\,^{\mathcal{M}}=&&\begin{array}{c} \dot{\mu} \\ m' \\ m \\ \mu \end{array}\left(\begin{array}{cccc}M_{\dot{\mu}}\,^{\dot{\nu}},D\delta_{\dot{\mu}}\,^{\dot{\nu}} \;\;& \bar{S}_{\dot{\mu}}\,^{n'} & \bar{S}_{\dot{\mu}}\,^n & K_{\dot{\mu}}\,^{\nu} \\ \bar{Q}_{m'}\,^{\dot{\nu}} & T_{m'}\,^{n'} & T_{m'}\,^n & S_{m'}\,^{\nu}\\ \bar{Q}_{m}\,^{\dot{\nu}} & T_{m}\,^{n'} & T_m\,^n & S_m\,^\nu \\ P_{\mu} \,^{\dot{\nu}} & Q_\mu \,^{n'} & Q_{\mu}\,^{n} & M_{\mu}\,^{\nu},D\delta_{\mu}\,^{\nu}\end{array}\right)\,.
 \label{groupelement}
\end{eqnarray}
The indices $(\mu,\;\dot\mu,\; m,\; m')$ are all SU(2) indices, and can be raised or lowered by the SU(2) metric. The (local) isotropy group acts from the left on $\mathcal{Z}_{\mathcal{A}}\,^{\mathcal{M}}$. Rewriting the superspace coordinates as
 \begin{equation}
\nonumber\mathcal{Z}_{\mathcal{A}}\,^{\mathcal{M}}=\left(\begin{array}{cc}z_A\,^M & z_A\,^{M'} \\z_{A'}\,^M & z_{A'}\,^{M'}\end{array}\right)=\left(\begin{array}{cc}{\rm I} & v \\0 & {\rm I} \end{array}\right)\left(\begin{array}{cc}u & 0 \\0 & u'\end{array}\right)\left(\begin{array}{cc}{\rm I} & 0 \\w & {\rm I}\end{array}\right)\\
\nonumber=\left(\begin{array}{cc}u+vu'w & vu' \\u'w & u'\end{array}\right)\,, 
\end{equation}
the half-coset space U(2,2$|$4)/U(1,1$|$2)$^2_+$  then corresponds to using $G_-=\left(\bar{S}_{\dot{\mu}}\,^n,\; K_{\dot{\mu}}\,^{\nu},T_{m'}\,^n, S_{m'}\,^{\nu}\right)$ and $G_0=\left(M,D,\bar{S}_{\dot{\mu}}\,^{n'} ,T_m\,^n\right.$ $\left.S_m\,^\nu,Q_{\mu}\,^{n},\bar{Q}_{m'}\,^{\dot{\nu}}, T_{m'}\,^{n'} \right)$ to gauge away $v,u,u'$, and one is left with the coordinates $w$.

More explicitly, the half-coset coordinates $w$ are given by: 
\begin{equation}
w_{M'}\,^{N}\equiv\left( z_{A'}\,^{M'}\right)^{-1}z_{A'}\,^{N}=\left( x_{\dot\mu}\,^\nu,\theta_{m'}\,^\nu,\bar{\theta}_{\dot\mu}\,^{n},y_{m'}^n\right)\,.
\label{projectivespace}
\end{equation}
Note that since the coordinate $w$ is defined in terms of ``ratios" of $z$ coordinates, which linearly realizes the global superconformal group $G$, the $w$ coordinates transforms as ratios of the transformation generated by the subgroups of $G$~\cite{Hatsuda:2007wr}. Explicitly, one can show that under a general global SU(2,2$|$4) element 
\begin{equation}
G_{\mathcal{M}}\,^{\mathcal{N}}=\left(\begin{array}{cc}a & b \\c & d\end{array}\right)
\end{equation}
the $w$ coordinates transform as 
\begin{equation}
w\rightarrow\left(wc+d\right)^{-1}\left(wa+b\right)\,.
\end{equation}
This non-linear realization of the superconformal group is why the coordinates $w$ are generally referred to as ``projective" superspace.

The superconformal generators, when written in these coordinates, take the form:
\begin{eqnarray}
\nonumber &&G_M\,^{N'}=\frac{\partial}{\partial w_{M'}\,^{N}}, \quad\quad\quad\quad\quad G_M\,^{N}=w_{P'}\,^N\frac{\partial}{\partial w_{P'}\,^{M}}\\
&&G_{M'}\,^{N'}=-w_{M'}\,^P\frac{\partial}{\partial w_{N'}\,^{P}},\quad\;\;G_{M'}\,^{N}=-w_{M'}\,^Pw_{Q'}\,^{N}\frac{\partial}{\partial w_{Q'}\,^{P}}\,.
\label{generators}
\end{eqnarray}
 The contraction of repeated indices will introduce extra minus sign if the R-symmetry index ($m,n'$) is contracted across odd number of R-symmetry index.\footnote{As a consequence, one has $\frac{\partial y_{m'}\,^n}{\partial y_{o'}\,^p}=-\delta_{n}\,^{p}\delta_{m'}\,^{o'}$ while $\frac{\partial x_{\dot\mu}\,^\nu}{\partial x_{\dot\rho}\,^\sigma}=\delta_{\nu}\,^{\sigma}\delta_{\dot\mu}\,^{\dot\rho}$.} For convenience, we label 
 \begin{eqnarray}
 G_w\equiv G_M\,^{N'}, \;\; G_u\equiv G_M\,^{N},\;\;G_{u'}=G_{M'}\,^{N'},\;\;G_v=G_{M'}\,^{N}
 \label{categorize}
 \end{eqnarray}
From this point of view, chiral superspace corresponds to the half coset U(2,2$|$4)/U(1,1$|$4)U(1,1)$_+$ where $G_w=(P_{\mu} \,^{\dot{\nu}},\; Q_\mu \,^{n'},\; Q_{\mu}\,^{n})$.

\section{From projective to on-shell $(\lambda,\tilde\lambda,\eta,\bar\eta)$ space.  \label{section3}}
Now that we've identified the dual space as projective superspace and obtained the form of the superconformal generators in this space, we can deduce how the generators act on the amplitudes by converting the projective coordinates to the on-shell coordinates, $(\lambda,\tilde\lambda,\eta,\bar\eta)$. Recall that in the original chiral superspace, the dual variables are related to the on-shell space via the hyper-plain constraints~\cite{DualConformal2}:
\begin{eqnarray}
\nonumber&&x_{i}^{\mu\dot\nu}-x_{i+1}^{\mu\dot\nu}=\lambda_i^\mu\tilde{\lambda}_i^{\dot\nu}\\
&&\theta_{i}^{I\mu}-\theta_{i+1}^{I\mu}=\lambda_i^\mu\eta_i^{I}
\end{eqnarray}
where $I=(1-4)$ are the SU(4) indices and $i$ labels the external legs of the color ordered amplitude. For projective superspace, the natural construction is to break the SU(4) down to SU(2)$\times$SU(2), with $\eta^{m'}\equiv\eta^I$ for $I=1,2$ and $\bar{\eta}_{m}\equiv\bar{\eta}_I$ for $I=3,4$. Thus the dual projective coordinates  $\left( x_{\dot\mu}\,^\nu,\theta_{m'}\,^\nu,\bar{\theta}_{\dot\mu}\,^{n},y_{m'}\,^n\right)$ are now related to the non-chiral on-shell space ($\lambda,\tilde\lambda, \eta, \bar{\eta}$) as :
\begin{eqnarray}
\nonumber x_{i\dot\mu}\,^\nu-x_{i+1\dot\mu}\,^\nu\;\;&=&\tilde{\lambda}_{i\dot\mu}\lambda_i\,^\nu\\
\nonumber\theta_{im'}\,^\nu-\theta_{i+1m'}\,^\nu\;&=&\eta_{im'}\lambda_i\,^\nu\\
\nonumber\bar{\theta}_{i\dot\mu}\,^{n}-\bar{\theta}_{i+1\dot\mu}\,^{n}\;\;&=&\tilde{\lambda}_{i\dot\nu}\bar{\eta}_{i}\,^n\\
y_{im'}\,^n-y_{i+1m'}\,^n&=&\eta_{im'}\bar\eta_i^n
\label{hyperplane}
\end{eqnarray}
Note that the R-coordinates are grassmann even. The action of the dual generators on the on-shell space can be derived by requiring that the generators are extended such that the above hyperplane constraints are maintained, i.e. the result of applying $\hat{G}_{\mathcal{M}}\,^{\mathcal{N}}\equiv \sum_{i=1}^nG_{i\mathcal{M}}\,^{\mathcal{N}}$ on the LHS of eqn.(\ref{hyperplane}) must match with the result of the extension terms acting on the RHS of eqn.(\ref{hyperplane}). From now on, hatted generators indicate that it is a complete sum of single site generators. We analyse each of the generators ($\hat{G}_w, \;\; \hat{G}_u,\;\;\hat{G}_{u'},\;\;\hat{G}_v$) separately. 
\subsection{Generators $\hat{G}_w$:}
The generators $\hat{G}_w=(\hat{P}_{\mu} \,^{\dot{\nu}},\; \hat{Q}_\mu \,^{n'}, \hat{\bar{Q}}_{m}\,^{\dot{\nu}}\;, \hat{T}_{m}\,^{n'} )$ are trivial in the on-shell space, since these corresponds to translations in $\left( x_{\dot\mu}\,^\nu,\theta_{m'}\,^\nu,\bar{\theta}_{\dot\mu}\,^{n},y_{m'}^n\right)$ and eq.(\ref{hyperplane}) is translational invariant. Thus these generator does not impose any constraint on the superamplitude. 

\subsection{Generators $\hat{G}_u,\hat{G}_{u'}$:}
The extension of the generators $\hat{G}_u$ and $\hat{G}_{u'}$ corresponds to the usual superconformal generators. To show this, we use $\hat{Q}_{\mu}\,^{n}$ as an example. From eq.(\ref{generators}) we have 
\begin{eqnarray}
\nonumber\hat{Q}_{\mu}\,^{n}&=&\sum_i w_{iP'}\,^n\frac{\partial}{\partial w_{iP'}\,^{\mu}}\\
&=&\sum_i -y_{im'}\,^n\frac{\partial}{\partial \theta_{im'}\,^{\mu}}+\bar{\theta}_{i\dot\mu}\,^n\frac{\partial}{\partial x_{i\dot\mu}\,^{\mu}}
\end{eqnarray}
To preserve eq.(\ref{hyperplane}), we see that one needs to modify 
\begin{eqnarray}
\hat{Q}^*_{\mu}\,^{n}=\hat{Q}_{\mu}\,^{n}+\sum_i \bar\eta_i^n\frac{\partial}{\partial\lambda^\mu_i}\,.
\end{eqnarray}
Thus one sees that when acting on the on-shell space, this is simply the original conformal susy generators $\hat{s}_\mu^n$. Following the conventions of~\cite{DualConformal2}, we denote dual generators by capital letters and small letters for ordinary superconformal generators. Similar result can be obtained for all other $\hat{G}_u$ and $\hat{G}_{u'}$ generators, which maps into original superconformal generators with the  role of susy and conformal susy  generators reversed, i.e. $\hat{S}\rightarrow \hat{q}, \hat{Q}\rightarrow \hat{s}$. Note that the dual generators $\hat{G}_u$ and $\hat{G}_{u'}$ corresponds to the original superconformal generators that are linear in ($\lambda,\tilde\lambda, \bar\eta^m,\eta_{m'}$) derivatives. Later we will use this fact to extend the dual conformal symmetry in six-dimensions to include half of the fermionic generators.

\subsection{Generators $\hat{G}_v$:}
The remaining 16 generators $\hat{G}_{v}$ are non-linear transformations in projective superspace. Here we will demonstrate that, once extended to the on-shell space, they become level-1 Yangian generators, $J^{(1)A}\equiv f^{ABC}J^{(0)B}J^{(0)C}$, where $J^{(0)}$ are the conventional superconformal generators labelled by $A$, and $f^{ABC}$ is the SU(2,2$|$4) structure constants. We analyse the $\hat{G}_{v}$s separately.
 \begin{itemize}
  \item 1. $\hat{K}_{\dot\mu}\,^{\nu}$:
  
  In projective space this takes the form 
  \begin{eqnarray}
 \nonumber \hat{K}_{\dot\mu}\,^{\nu}&=&\sum_{i=1}^n\left(\bar{\theta}_{i\dot\mu}\,^m\theta_{in'}\,^{\nu}\frac{\partial}{\partial y_{in'}\,^{m}} +\bar{\theta}_{i\dot\mu}\,^m x_{i\dot\nu}\,^{\nu}\frac{\partial}{\partial \bar{\theta}_{i\dot\nu}\,^{m}} +x_{i\dot\mu}\,^\rho \theta_{in'}\,^{\nu}\frac{\partial}{\partial \theta_{in'}\,^{\rho}} \right.\\
 &&\left.\quad\quad\quad\quad\quad+x_{i\dot\mu}\,^\rho x_{i\dot\nu}\,^{\nu}\frac{\partial}{\partial x_{i\dot\nu}\,^{\rho}}\right)\,.
   \end{eqnarray}
  To preserve the four constraint equations, one arrive at the following extension 
  \begin{eqnarray}
 \nonumber \hat{K}^*_{\dot\mu}\,^{\nu}&=&\hat{K}_{\dot\mu}\,^{\nu}+\sum_{i=1}^n\left(\frac{1}{2}\left(x_i+x_{i+1}\right)_{\dot\mu}\,^\rho\lambda_i^\nu\frac{\partial}{\partial \lambda_i^\rho}+\frac{1}{2}\left(x_{i}+x_{i+1}\right)_{\dot\rho}\,^\nu\tilde\lambda_{i\dot\mu}\frac{\partial}{\partial\tilde\lambda_{i\dot\rho}}\right.\\
&& \left.\quad\quad+\frac{1}{2}\left(\theta_{i}+\theta_{i+1}\right)_{m'}\,^\nu\tilde\lambda_{i\dot\mu}\frac{\partial}{\partial \eta_{im'}}+\frac{1}{2}\left(\bar{\theta}_{i}+\bar{\theta}_{i+1}\right)_{\dot\mu}\,^m\lambda_i^\nu\frac{\partial}{\partial \bar\eta_i^m}\right)\,.
\label{extension}
  \end{eqnarray}  
Note that this extension is different from the usual result derived from chiral superspace~\cite{DualConformal2}, it is manifestly symmetric in both $\lambda$ and $\tilde\lambda$. This implies that the conformal inversion rules for $\lambda$ and $\tilde\lambda$ will be symmetric as well. The tree-level amplitudes are invariant under the shifted dual conformal boost generator as~\cite{DualConformal2}: 
\begin{equation}
\left(\hat{K}^*_{\dot\mu}\,^{\nu}+\sum_{i=1}^nx_{i\dot\mu}\,^{\nu}\right)\mathcal{A}_n=0
\label{shift}
\end{equation}
While the above result was derived for the amplitudes expressed in chiral superspace, it simply follows from 
\begin{equation}
I\left[\mathcal{A}_n\right]=\left(\prod_{i=1}^nx_i^2\right)\mathcal{A}_n,\;\;{\rm and}\;\;IPI=K
\end{equation}
which holds for projective superspace as well. 

Following~\cite{Drummond:2009fd}, one can rewrite the extension terms in eq.(\ref{extension}) as bi-local products of the original superconformal generators. Using 
\begin{eqnarray}
\nonumber x_{i\dot\mu}\,^\nu&=&x_1-\sum_{j=1}^{i-1}\lambda_j^\nu\tilde{\lambda}_{j\dot\mu},\quad\quad\theta_{im'}^\nu=\theta_{1m'}^\nu-\sum_{j=1}^{i-1}\lambda_j^\nu\bar{\eta}_{jm'}\\
\nonumber\bar{\theta}_{i}\,^{\dot\mu n}&=&\bar{\theta}_{1}^{\dot\mu n}-\sum_{j=1}^{i-1}\tilde{\lambda}_j^{\dot\nu}\eta_{j}^n,\quad\quad y_{im'}^n=y_{1m'}^n-\sum_{j=1}^{i-1}\eta_{jm'}\bar\eta_j^n
\end{eqnarray}
and denoting $\hat{K}^*_{\dot\mu}\,^{\nu}+\sum_{i=1}^nx_{i\dot\mu}\,^{\nu}=\hat{K}_{\dot\mu}\,^{\nu}+\Delta \hat{K}_{\dot\mu}\,^{\nu}$, one sees that the $(x_1,\theta_1,\bar\theta_1)$ dependent terms in $\Delta \hat{K}_{\dot\mu}\,^{\nu}$ can be written as  
 \begin{eqnarray}
 \nonumber&\sum_{i=1}^n&\left(x_{1\dot\mu}\,^\rho\lambda_i^\nu\frac{\partial}{\partial \lambda_i^\rho}+x_{1\dot\rho}\,^\nu\tilde\lambda_{i\dot\mu}\frac{\partial}{\partial\tilde\lambda_{i\dot\rho}}+\theta_{1m'}\,^\nu\tilde\lambda_{i\dot\mu}\frac{\partial}{\partial \eta_{im'}}+\bar{\theta}_{1\dot\mu}\,^m\lambda_i^\nu\frac{\partial}{\partial \bar\eta_i^m}+x_{1\dot\mu}\,^{\nu}\right)\\
 &=&\left(x_{1\dot\mu}\,^\rho m^\nu\,_{\rho}+x_{1\dot\rho}\,^\nu\tilde{m}_{\dot\mu}\,^{\dot\rho}+x_{1\dot\mu}\,^{\nu}d+\theta_{1m'}\,^\nu\bar{q}_{\dot\mu}\,^{m'}+\bar{\theta}_{1\dot\mu}\,^m q^\nu_m\right)\,.
  \end{eqnarray}  
where the generators ($q,\bar{q},d,\tilde{m},m$) are part of the original superconformal generators, and hence vanishes trivially on the amplitude. The remaining terms in $\Delta \hat{K}_{\dot\mu}\,^{\nu}$ takes the form 
 \begin{eqnarray}
 \nonumber &&-\sum_{i=1}^n\sum_{j<i}^n\left(p_{j\dot\mu}\,^\rho m_i^\nu\,_{\rho}+p_{j\dot\rho}\,^\nu\tilde{m}_{i\dot\mu}\,^{\dot\rho}+q_{jm'}\,^\nu\bar{q}_{i\dot\mu}\,^{m'}+\bar{q}_{j\dot\mu}\,^mq^\nu_{im}+p_{j\dot\mu}\,^\nu d_i\right)\\
\nonumber &&-\frac{1}{2}\sum_{i=1}^n\left(p_{i\dot\mu}\,^\rho m_i^\nu\,_{\rho}+p_{i\dot\rho}\,^\nu\tilde{m}_{i\dot\mu}\,^{\dot\rho}+q_{im'}\,^\nu\bar{q}_{i\dot\mu}\,^{m'}+\bar{q}_{i\dot\mu}\,^mq^\nu_{im}+p_{i\dot\mu}\,^\nu d_i\right)+\frac{1}{2}p_{\dot\mu}\,^\nu\,.\\
  \end{eqnarray}  
 We can add to the above the following term which trivially vanishes on the amplitude,
 \begin{eqnarray}
\nonumber\frac{1}{2}\sum_{j=1}^n\left(p_{j\dot\mu}\,^{\rho} m^{\nu}\,_{\rho}+p_{j\dot\rho}\,^{\nu}\tilde{m}_{\dot\mu}\,^{\dot\rho}+q_{jm'}\,^{\nu}\bar{q}_{\dot\mu}\,^{m'}+\bar{q}_{j\dot\mu}\,^m q^\nu_{m}+p_{j\dot\mu}\,^\nu d\right)-\frac{1}{2}p_{\dot\mu}\,^\nu\,,\\
\end{eqnarray}  
we finally arrive at the form  
\begin{eqnarray}
 \nonumber \Delta \hat{K}_{\dot\mu}\,^{\nu}=-\frac{1}{2}\sum_{j<i}^n\left(p_{j\dot\mu}\,^\rho m_i^\nu\,_{\rho}+p_{j\dot\rho}\,^\nu\tilde{m}_{i\dot\mu}\,^{\dot\rho}+q_{jm'}\,^\nu\bar{q}_{i\dot\mu}\,^{m'}+\bar{q}_{j\dot\mu}\,^mq^\nu_{im}+p_{j\dot\mu}\,^\nu d_i-(i\leftrightarrow j)\right)\\
\end{eqnarray}  
which is the standard level one generator $J^{(1)}_{\dot\mu}\,^{\nu}$ although in a non-chiral form.

   \item 2. $\hat{S}_{\dot\mu}\,^{n},\;\hat{S}_{m'}\,^{\nu}$
   
   We now look at the generator $\hat{S}_{\dot\mu}\,^{n}$ ($\hat{S}_{m'}\,^{\nu}$ can be derived in the same fashion). We have: 
   \begin{eqnarray}
  \nonumber \hat{S}_{\dot\mu}\,^{n}&=&\sum_{i=1}^n\left(-\bar{\theta}_{i\dot\mu}\,^p \bar{\theta}_{i\dot\nu}\,^{n}\frac{\partial}{\partial \bar{\theta}_{i\dot\nu}\,^{p}}+\bar{\theta}_{i\dot\mu}\,^p y_{im'}\,^{n}\frac{\partial}{\partial y_{im'}\,^{p}}+x_{i\dot\mu}\,^\nu \bar{\theta}_{i\dot\nu}\,^{n}\frac{\partial}{\partial x_{i\dot\nu}\,^{\nu}}\right.\\
  &&\left.\quad\quad\quad-x_{i\dot\mu}\,^\nu y_{im'}\,^{n}\frac{\partial}{\partial \theta_{im'}\,^{\nu}}\right)\,.
  \end{eqnarray}
   To preserve the constraint equations we extend it by 
    \begin{eqnarray}
  \nonumber \hat{S}_{\dot\mu}\,^{*n}&=&\hat{S}_{\dot\mu}\,^{n}+\sum_{i=1}^n\left(\frac{1}{2}(\bar{\theta}_i+\bar{\theta}_{i+1})_{\dot\alpha}\,^{n}\tilde\lambda_{i\dot\mu}\frac{\partial}{\partial\tilde\lambda_{i\dot\alpha}}+\frac{1}{2}\left(x_i+x_{i+1}\right)_{\dot\mu}\,^\alpha\bar\eta_i^n\frac{\partial}{\partial\lambda_i^\alpha}\right.\\
  &&\left.\quad\quad\quad-\frac{1}{2}(y_{i}+y_{i+1})_{m'}\,^{n}\tilde\lambda_{i\dot\mu}\frac{\partial}{\partial\eta_{im'}}-\frac{1}{2}(\bar{\theta}_i+\bar{\theta}_{i+1})_{\dot\mu}^p\bar\eta^n_i\frac{\partial}{\partial \bar\eta_i^p}\right)\,.
  \end{eqnarray}
Since the dual generators $(Q_{\mu}\,^n,\bar Q_{\dot\mu}\,^{n'})$ just correspond to the conformal susy generators $(s_{\mu}\,^n,\bar s_{\dot\mu}\,^{n'})$ which vanishes on the amplitude, using the algebra $[K_{\dot\nu}\,^\mu, Q_{\rho}\,^n]\sim \delta_\rho^\mu S^n_{\dot\nu}$ along with eq.(\ref{shift}), one can deduce: 
   \begin{equation}
\left(\hat{S}_{\dot\mu}\,^{*n}+\sum_{i=1}^n\bar{\theta}_{i\dot\mu}\,^{n}\right)\mathcal{A}_n=0
\label{Fermshift}
\end{equation}
and similar result for $\hat{S}_{m'}\,^{*\nu}$. Following similar steps as above, we find the extended part 
\begin{eqnarray}
 \nonumber \Delta \hat{S}_{\dot\mu}\,^{n}=-\frac{1}{2}\sum_{j<i}^n\left(\bar{q}_{j\dot\nu}\,^n \tilde{m}_{i\dot\mu}\,^{\dot\nu}+\frac{1}{2}\bar{q}_{j\dot\mu}\,^n(d+c)_j+p_{j\dot\mu}\,^\alpha s_{i\alpha}\,^{n}-r_{jm'}\,^n\bar{q}_{i\dot\mu}\,^{m'}-\bar{q}_{j\dot\mu}\,^pr^n_{ip}-(i\leftrightarrow j)\right)\\
\end{eqnarray}  
   which again matches with that of $J^{(1)}_{\dot\mu}\,^{n}$.

  \item 4. $\hat{T}_{m'}\,^{n}$

 Non-trivial dual R-symmetry generators is a generic feature in theories whose on-shell space is non-chiral. A prime example is the three-dimensional $\mathcal{N}=6$ super-Chern Simons theory~\cite{ABJM}. Here we have: 
  \begin{eqnarray}
  \nonumber\hat{T}_{m'}\,^{n}&=&-\sum_{i=1}^n\left(-\theta_{im'}\,^\mu y_{ip'}\,^{n}\frac{\partial}{\partial \theta_{ip'}\,^{\mu}}+\theta_{im'}\,^\mu \bar{\theta}_{i\dot\mu}\,^{n}\frac{\partial}{\partial x_{i\dot\mu}\,^{\mu}}+y_{im'}\,^m y_{ip'}\,^{n}\frac{\partial}{\partial y_{ip'}\,^{m}}\right.\\
 && \quad\quad\left.-y_{im'}\,^m\bar{\theta}_{i\dot\mu}\,^{n}\frac{\partial}{\partial \bar{\theta}_{i\dot\mu}\,^{m}}\right)\\
  \end{eqnarray}
  The extension to the on shell space is given by 
  \begin{eqnarray}
  \nonumber\hat{T}_{m'}\,^{*n}&=&\hat{T}_{m'}\,^{n}+\sum_{i=1}^n\left(\frac{1}{2}(\theta_i+\theta_{i+1})_{m'}\,^{\mu}\bar\eta_i^n\frac{\partial}{\partial\lambda_i^\mu}-\frac{1}{2}(\bar{\theta}_i+\bar{\theta}_{i+1})_{\dot\mu}\,^n\eta_{im'}\frac{\partial}{\partial\tilde{\lambda}_{i\dot\mu}}\right.\\
  &&\left.\quad\quad-\frac{1}{2}(y_i+y_{i+1})_{p'}\,^n\eta_{im'}\frac{\partial}{\partial\eta_{ip'}}-\frac{1}{2}(y_i+y_{i+1})_{m'}\,^p\bar\eta_i^{n}\frac{\partial}{\partial\bar\eta_{i}^{p}}\right)
  \end{eqnarray}
 Using the algebra $\{Q_{\mu}\,^n, S_{m'}\,^\nu\}\sim \delta_\mu^\nu T_{m'}\,^n$, one can deduce, from eq.(\ref{Fermshift}),
 \begin{equation}
 \left( \hat{T}_{m'}\,^{p*}+\sum_{i=1}^ny_{im'}\,^p\right)\mathcal{A}_n=0\,.
  \label{yshift}
 \end{equation} 
This allows us to arrive at the final bi-local form 
  \begin{eqnarray}
  \Delta\hat{T}_{m'}\,^{n}=\sum_{j<i}\left(q_{jm'}\,^{\mu}s_{i\mu}^n-\bar{q}_{j\dot\mu}\,^n\bar{s}_{im'}^{\dot\mu}+r_{jp'}\,^nr_{im'}\,^{p'}+r_{jm'}\,^pr_{ip}^{n}\right)\,,
  \label{T}
  \end{eqnarray}
  which indeed correspond to that of $J^{(1)}_{m'}\,^{n}$. 
\end{itemize}

Thus in conclusion, the DSC invariance in projective superspace implies 16 non-trivial level-1 Yangian like constraints on the on-shell space, $(J^{(1)}_{\dot\mu}\,^{\nu}, \,J^{(1)}_{\dot\mu}\,^{n},\, J^{(1)}_{m'}\,^{\nu},\,J^{(1)}_{m'}\,^{n})$. The remaining 48 Yangian generators can be obtained by (anti)commuting the original superconformal generators, those that do not correspond to $\hat{G}_u,\hat{G}_{u'}$, with $\hat{G}_w$.
\begin{figure}
\begin{center}
\includegraphics[scale=1.1]{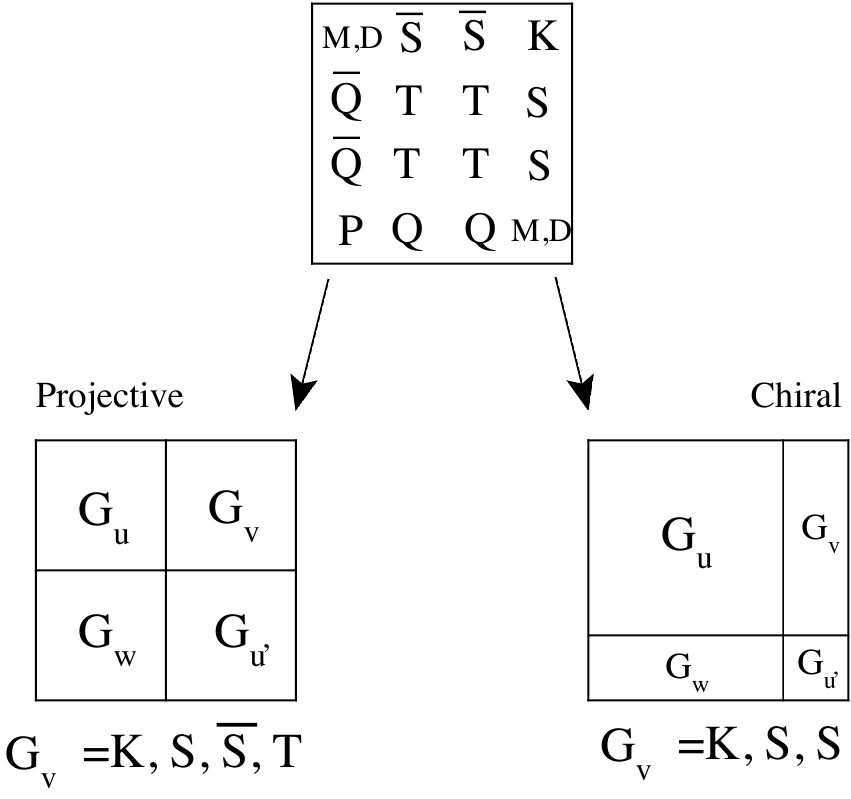}
\caption{The segregation of the DSC generators into level-0 and level-1 Yangian generators in the framework of projective and chiral superspace. The level-1 generators corresponds to the dual $\hat{G}_v$ generators. }
\label{supergroup}
\end{center}
\end{figure}

\begin{figure}
\begin{center}
\includegraphics[scale=0.9]{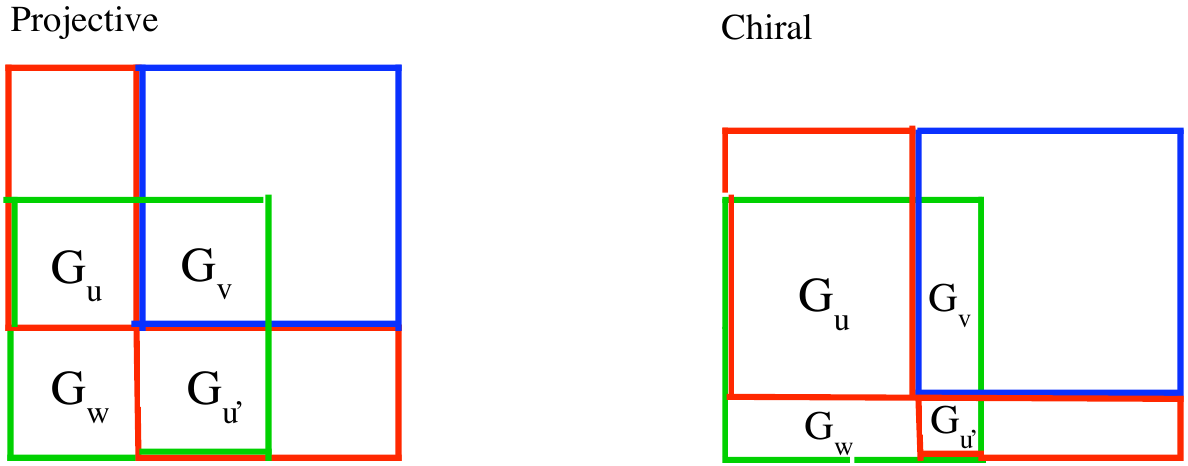}
\caption{Extension from the DSC generators to the full Yangian including level-0 and level-1 generators. The green contour encircles the DSC generators, the blue encircles the level-1 Yangian generators and the red contour encircles the level-0 or the original superconformal generators.}
\label{fundamental}
\end{center}
\end{figure}

In the half-coset description, it is clear which dual generators corresponds to ordinary superconformal generators, ($\hat{G}_u,\hat{G}_{u'}$), and which to level-1 Yangian generators ($\hat{G}_v$). This identification is valid for chiral superspace as well if one generalize the half-coset. Take a general half-coset of the U(2,2$|$4) group, U(2,2$|$4)/U(1,1$|$4-n)U(1,1$|$n)$_+$, we see that  $n=0$ corresponds to chiral and $n=2$ to projective superspace. Thus we see that for both cases, the isotropy group generators of the half-coset corresponds to the Yangian generators, where the ($\hat{G}_u,\hat{G}_{u'}$) are the level-0 and ($\hat{G}_v$) are the level-1 generators. The remaining half-coset generators ($\hat{G}_w$) are the trivial generators. This structure is shown in fig.\ref{supergroup}. This construction can be further extended to include the remaining Yangian generators, which has the level-1 and the level-0 ordinary superconformal generators that are not included in the DSC generators. This extension is schematically given in fig.\ref{fundamental}, where the green contour encircles the DSC generators, the blue encircles the level-1 Yangian generators and the red encircles the ordinary superconformal generators. The generators in the extended contours can be identified by their index structures, which are simply periodic repeats of those inside the green contour and given in eq.(\ref{groupelement}). 

Finally, we comment on the similarity between $T_{m'}\,^n$ and the recent bonus U(1) Yangian generator $\mathfrak{B}$ found in~\cite{Beisert:2011pn} which schematically take the form: 
\begin{equation}
\mathfrak{B}=\sum_{k=1}^{n-1}\sum_{j=k+1}^n\left(q_ks_j-\bar q_k\bar s_j-(i\leftrightarrow j)\right)\,.
\label{B}
\end{equation}
Although $T_{m'}\,^n$  in eq.(\ref{T}) is similar to the above, they are not equivalent. This can be seen by noting that combining $\mathfrak{B}$ with the ordinary susy and conformal susy generators generate the known dual superconformal generators:
\begin{equation}
[\mathfrak{B},s]=-\hat{S},\;[\mathfrak{B},q]=\hat{Q}
\end{equation}
On the other hand $T_{m'}\,^n$ is simply the dual $R$-symmetry generator and is part of the original SU(2,2$|$4) Yangian generator, thus it is implied by the other level-1 and level-0 generators. The difference can also be seen from the fact that the supercharges transform with the same weight:
\begin{equation}
[T_{m'}\,^n ,s]=\hat{S},\;[T_{m'}\,^n,q]=\hat{Q}\,.
\end{equation}
One can extend DSC symmetry from SU(2,2$|$4) to U(2,2$|$4), however the extra dual U(1) charge simply correspond to the original helicity generator since the U(1) is a combination of $G_u,G_{u'}$. Thus the bonus generator $\mathfrak{B}$ does not correspond to the U(1) extension of DSC symmetry.

\subsection{Non-chiral DSC symmetry as a descendant from six dimensions}
While it is expected that the above non-chiral DSC symmetry is a symmetry of the amplitudes, since it is merely a different representation of the same algebra, it is instructive to take the view that it is a descendant symmetry from six dimensions. From this point of view, the six-dimensional dual conformal symmetry is inherited by the four-dimensional non-chiral amplitude. The superconformal invariance of the four-dimensional massless theory then enhances the dual symmetry to the full DSC symmetry. On the other hand, since the six-dimensional theory and its four-dimensional massive descendant is only super Poincar\'{e} invariant, the dual conformal symmetry can only be enlarged to half of the fermionic generators of the full DSC symmetry. Thus the four-dimensional massless theory enjoys a symmetry enhancement both in terms of the original and dual symmetry. We show this by explicit reduction of the six-dimensional on-shell variables.

Amplitudes in six-dimensions can be completely written in terms of (super)momenta, $(\mathfrak{p}_i^{AB}, \mathfrak{q}^A_{ia},\tilde{\mathfrak{q}}_{iA\dot a})$, where $A,B$ are the six dimensional SU$^*$(4) Lorentz indices, and $a,\dot a$ are the SO(4)$\sim$ SU(2)$\times$SU(2) little group indices.\footnote{A vector in six dimensions is in the anti-symmetric representation of SU$^*$(4). The $^*$ represents that it is a pseudoreal representation. For details of the six dimensional on-shell superspace see~\cite{DHS}} Note that there are two types of supermomenta, $\mathfrak{q}_i$ and $\tilde{\mathfrak{q}}_i$, one for each SU(2) of the $\mathcal{N}=(1,1)$ R symmetry group SU(2)$\times$SU(2). It was shown~\cite{Dennen:2010dh} that under dual conformal inversion, the six-dimensional tree amplitude inverts as 
\begin{equation}
I\left[\mathcal{A}_n^{D=6}\right]=(x_1^2)^2\left(\prod_{i=1}^n x_i^2\right)\mathcal{A}_n^{D=6}\,.
\end{equation} 
This is exactly the same form as the four dimensional result except for the extra factor of $(x_1^2)^2$. The origin of this discrepancy is due to the mismatch of the (mass)dimensions between the momentum delta function $\delta^6(P)$ and supermomentum delta function $\delta^8(Q)$. To make the connection to four dimensions more straightforward, one defines a function $f_n$ as
\begin{equation}
\mathcal{A}_n^{D=6}=\delta^6(P)\delta^4(Q)\delta^4(\tilde{Q})f_n^{D=6}\,.
\end{equation}
The usual $\delta^8(Q)$ is now written as $\delta^4(Q)\delta^4(\tilde{Q})$ due to the non-chiral $\mathcal{N}=(1,1)$ symmetry of the six dimensional SYM. We then have 
\begin{equation}
I\left[f_n^{D=6}\right]=\left(\prod_{i=1}^n x_i^2\right)f_n^{D=6}\,.
\label{invert6}
\end{equation}
If one restricts the six-dimensional kinematics to a four-dimensional subplane, the on-shell variables reduces straightforwardly to four dimensional ones, 
\begin{eqnarray}
\mathfrak{p}_i^{AB}\rightarrow
	\left(\begin{array}{cc}
			0
			&p_{i\alpha}\,^{\dot\beta}    \\
			-p_i^{\dot\alpha}\,_\beta &
			0
		\end{array}\right)\,, \qquad \mathfrak{q}^A_{i}\rightarrow\left(\begin{array}{c}
			q_{i\alpha}^1\\ 
			-\bar{q}_i\,^{\dot\alpha}_4
		\end{array}\right)\,,\qquad 
	\tilde{\mathfrak{q}}_{iA}\rightarrow\left(\begin{array}{c} q_i^{\alpha2}\\ 
		-\bar{q}_{i\dot\alpha3}
		\end{array}\right)\,.\nn\\
		\label{reduction}
\end{eqnarray}
Note that the six-dimensional chiral supermomenta reduces to both chiral and anti-chiral supermomenta in four dimensions. The four-dimensional R-index is not fixed, and we've chosen to have $(I=1,2)$ to be chiral and $(I=3,4)$ to be anti-chiral so one can straightforwardly match with the projective superspace result. The result is a non-chiral $f_n^{D=4}$, which inverts in the same way as the traditional chiral version due to eq.(\ref{invert6}). The covariance under the non-chiral conformal inversion then implies covariance under non-chiral conformal boost. One can also directly match the six dimensional dual generators with the four dimensional non-chiral version. For example, the dual conformal boost generator in six dimensions takes the form~\cite{Dennen:2010dh}:
\begin{eqnarray}
D=6:\hat{K}^{\mu}&=&\sum_{i}\left[
\left(2 \, x^\mu_i x^\nu_i-x_i^2 \, \eta^{\mu\nu}\right)\frac{\partial}{\partial x_{i}^{\nu}}
+\theta^A_i(\sigma^{\mu})_{AB}x_i^{BC}\frac{\partial}{\partial \theta^C_i}
+\tilde{\theta}_{iA}(\tilde{\sigma}^{\mu})^{AB}x_{iBC}\frac{\partial}{\partial \tilde{\theta}_{iC}}\right] \nn\\
&+&\frac{1}{2} \left[\lambda^{Aa}_{i}(\sigma^\mu)_{AB}(x_i+x_{i+1})^{BC}\frac{\partial}{\partial \lambda^{Ca}_{i}}
-(\theta_i+\theta_{i+1})^A(\sigma^\mu)_{AB}\lambda_{ia}^{B}\frac{\partial}{\partial \eta_{ia}} \right.\nn\\
&&+\left.\tilde{\lambda}_{iA\da}(\tilde{\sigma}^\mu)^{AB}(x_i+x_{i+1})_{BC}\frac{\partial}{\partial \tilde{\lambda}_{iC\da}}
-(\tilde{\theta}_i+\tilde{\theta}_{i+1})_{A}(\tilde{\sigma}^{\mu})^{AB}\tilde{\lambda}^{\da}_{iB}\frac{\partial}{\partial \tilde{\eta}_{i}^{\da}}\right] \,,\nn\\
\end{eqnarray}
which is indeed the same as eq.(\ref{extension}) if one translate the six dimensional $x,\theta,\tilde\theta$ to four dimensional ones as in eq.(\ref{reduction}) and identify the on-shell variables as follows:
\begin{eqnarray}
&&	\lambda^A{}_a=\left(\begin{array}{cc}
			0&\lambda_\alpha\\ 
			\tilde\lambda^{\dot\alpha}&0
		\end{array}\right)\,,\quad 
	\tilde \lambda_{A\dot a}=\left(\begin{array}{cc}
		0  &\lambda^\alpha\\ 
		-\tilde\lambda_{\dot\alpha}&0 
		\end{array}\right)\,,\quad \eta_{ai}=\left(
			\eta^1,\;
			\bar\eta_4\right)\,,\; 
	\tilde \eta^{\dot a}=\left( \bar{\eta}_{3},\;\eta^{2}\right)\,.
\end{eqnarray}
Once we've established the dual conformal covariance under the generator  $\hat{K}^*_{\dot\mu}\,^{\nu}$, the covariance under $\hat{S}^*_{\dot\mu}\,^{n}$ and $\hat{S}^*_{m'}\,^{\nu}$ can be established by noting that the amplitudes are invariant under $\hat{Q}_\mu\,^n=\hat{s}_\mu\,^n$ and $\hat{\bar{Q}}_{\dot\mu}\,^{n'}=\hat{\bar{s}}_{\dot\mu}\,^{n'}$, which is the usual conformal susy generators. Similarly the covariance under $\hat{T}^*_{m'}\,^n$ can be established. Thus one sees that the superconformal invariance of the massless amplitude enhances the dual conformal symmetry to the full DSC symmetry.

As a side note, the six dimensional parent amplitude actually enjoys ``half" of the dual {\it super} conformal symmetry, albeit trivially.(Readers only interested in the four dimensional story can skip to the next subsection). We first discuss the six dimensional on-shell space in slightly more detail. The R-symmetry of $\mathcal{N}=(1,1)$ supersymmetry is SU(2)$\times$SU(2). The fermionic variables of this on-shell space is $(\eta_a,\tilde{\eta}_{\dot a})$, which carries the little group index. The $\eta, \tilde{\eta}$s can be understood as positive eigensates of the $J_z$ generator of the two SU(2)s, i.e. $[J_z, \eta^+_a]=\eta^+_a,\; [\tilde{J}_z, \tilde{\eta}^+_{\dot{a}}]=\tilde{\eta}^+_{\dot{a}}$. In this notation, the fermionic generators can be written as 
\begin{equation}
(\hat{\mathfrak{s}}^{A\pm},\;\hat{\mathfrak{q}}^{A\pm},\;\hat{\tilde{\mathfrak{s}}}_{A}^{\pm},\;\hat{\tilde{\mathfrak{q}}}_{A}^{\pm})\,.
\end{equation}
The function $f_n^{D=6}$ is invariant under $\hat{\mathfrak{q}}^{A\pm},\;\hat{\tilde{\mathfrak{q}}}_{A}\,^{\pm}$, but not $\hat{\mathfrak{s}}^{A\pm},\;\hat{\tilde{\mathfrak{s}}}_{A}^{\pm}$. In terms of the original on-shell variables, ($\lambda^A_a,\eta^+_a,\tilde\lambda_{A\dot a},\tilde\eta^+_{\dot a}$), generators $\hat{\mathfrak{q}}^{A-}, \hat{\tilde{\mathfrak{q}}}^{-}_A$ are linear in derivatives  while $\hat{\mathfrak{q}}^{A+}, \hat{\tilde{\mathfrak{q}}}^{+}_A$ are independent of derivatives. From our previous discussion we see that ordinary superconformal generators that are linear in derivatives correspond to dual generators with the identification $\hat{\mathfrak{q}}\rightarrow \hat{S}$ and $\hat{\mathfrak{s}}\rightarrow \hat{Q}$. Thus we see that the amplitudes are invariant under dual $\hat{S}^{A-}, \hat{\tilde{S}}^{-}_A$ generators. Finally the six dimensional hyper plane constraint 
\begin{equation}
\theta^{+A}_i-\theta^{+A}_{i+1}=\lambda^{Aa}_{i}\eta^+_{ia}, \qquad\tilde\theta^{+A}_i-\tilde\theta^{+A}_{i+1}=\tilde\lambda_{iA\dot{a}}\tilde\eta^{+\dot a}_{i}
\label{sixd}
\end{equation}
also implies the trivial invariance of the amplitude under $\hat{Q}^{-}_A=\sum^n_{i=1}\frac{\partial}{\partial \theta^{+A}_i}$, and $\hat{\tilde{Q}}^{A-}=\sum^n_{i=1}\frac{\partial}{\partial \tilde\theta^{+}_{iA}}$.

Thus in conclusion, we find that the dual conformal symmetry of $f^{D=6}_n$ can be extended to the following supercharges
\begin{equation}
\left(\hat P,\hat M,\hat D, \hat K\right)\rightarrow\left(\hat P,\hat M,\hat D,\hat K\right)+\left(\hat Q^-,\hat S^-\right)+\left(\hat{\tilde{Q}}^-, \hat{\tilde{S}}^-\right)\,.
\label{6Dextension}
\end{equation}
The charges form a subgroup of two six-dimensional superconformal group OSp$^*$(8$|$2) with opposite chirality. The fermionic generators transform under the supergroup as 
\begin{equation}
(1,0)^{-}\oplus(0,1)^{-}
\end{equation}   
where the $-$ indicate their charges under the two U(1)s of the Sp$^*$(2)$\sim$SU(2) R-symmetry and the entry in the parenthesis indicate their chirality within the SO$^*$(8). We stress, however, the extension does not imply new constraints outside of the dual conformal and original supersymmetry. Thus the extension of the dual symmetry is a trivial extension. As a consequence the four-dimensional massive amplitudes enjoys half of the fermionic symmetries as well. 

\section{Non-chiral amplitudes in dual superspace\label{section4}}
Given that the amplitudes enjoy the non-chiral DSC symmetry, it is natural to write them in terms of dual superspace coordinates. The conventional chiral representation, 
\begin{equation}
\mathcal{A}_n^{N^kMHV}= \mathcal{A}_n^{MHV}\times \mathcal{P}_n^{N^kMHV}
\label{Define1}
\end{equation} 
is no longer valid since the fermionic dependence of MHV amplitudes is no longer simply supermomentum delta function. The simplest choice is then to use the six dimensional analogue, 
\begin{equation}
\mathcal{A}_n=\delta^4(x_1-x_{n+1})\delta^4(\theta_1-\theta_{n+1})\delta^4(\bar{\theta}_1-\bar{\theta}_{n+1})f^{D=4}_n
\label{Define2}
\end{equation} 
where the three delta functions enforces momentum and supermomentum conservation. The function $f^{D=4}_n$ is of fermionic degree $2(n-4)$ in $\theta_{m'}\,^\mu,\bar{\theta}_{\dot\mu}\,^m$. This counting of fermionic degrees of freedom can be derived easily from six dimensions. Alternatively, one can obtain this conclusion by Fourier transforming half of the $\eta$s in the usual chiral representation. As mentioned previously, the function $f^{D=4}_n$ inverts with equal weight on all external lines:
\begin{equation}
I\left[f^{D=4}_n\right]=\left(\prod_{i=1}^n x_i^2\right)f^{D=4}_n.
\label{invert}
\end{equation}
At four-point the one has~\cite{Hatsuda:2008pm}
\begin{equation}
f^{D=4}_4=\frac{1}{x^2_{13}x^2_{24}}.
\end{equation}

Since MHV amplitudes are DSC invariant, the chiral separation in eq.(\ref{Define1}) defines $\mathcal{P}_n^{N^kMHV}$ that satisfies the full dual symmetry. The definition of $f^{D=4}_n$ in eq.(\ref{Define2}), however, does not enjoy the full DSC symmetry. To see this one notes that the function $\delta^4(x_1-x_{n+1})\delta^4(\theta_1-\theta_{n+1})\delta^4(\bar{\theta}_1-\bar{\theta}_{n+1})$, which in on-shell space takes the form $\delta^4(\sum_i \lambda^\mu_i\tilde\lambda_{i\dot\mu})\delta^4(\sum_i \lambda_{i\mu}\bar\eta_{im'})$         $\;$
$\delta^4(\sum_i \tilde\lambda_{i\dot\mu}\eta^{m}_i)$, by itself is not invariant under 
\begin{equation}
\left(\bar{s}_{m'}\,^{\dot\nu},\,\hat{s}_{\mu}\,^{n},\,\hat{t}_{m}\,^n, \,\hat{t}_{m'}\,^{n'}\right).
\end{equation}
This translates to broken symmetry under the following DSC generators 
\begin{equation}
\left(\hat{\bar{Q}}_{m'}\,^{\dot\nu},\,\hat{Q}_{\mu}\,^{n},\,\hat{T}_{m}\,^n, \,\hat{T}_{m'}\,^{n'}, \hat{\bar{S}}_{\dot\mu}\,^{n},\,\hat{S}_{m'}\,^{\mu}\right).
\end{equation}
Since the entire amplitude enjoys the full DSC symmetry, $f^{D=4}_n$ must be anomalous under these symmetries in order to compensate the anomalies coming from the (super)momentum delta functions. One thus conclude that the function $f^{D=4}_n$ is only invariant under 
\begin{equation}
\left(\hat{G}_w,\;\hat{D},\;\hat{M}^\alpha\,_\beta,\hat{\bar{M}}^{\dot\alpha}\,_{\dot\beta}, \;\hat{\bar{S}}_{m'}\,^{\dot\nu},\,\hat{S}_{\mu}\,^{n},\,\hat{K}^*_{\dot\mu}\,^\nu\right).
\label{unbrokenG}
\end{equation}
We will come back to the possibility of having a full DSC covariant representation at the end of this section. We first discuss the relationship between four-dimensional massless non-chiral amplitudes and their six-dimensional parent amplitudes. We will then construct $f_5^{D=4}$ which would be the first dual conformal covariant function with fermionic dependency. This function plays the analogous role of the ``R"-function in the chiral representation. 
\subsection{Four-dimensional non-chiral amplitudes$\rightarrow$ massive amplitudes}
 In this subsection, we will give a procedure to obtain the six-dimensional amplitude from the non-chiral four-dimensional massless amplitude. 
As mentioned previously, the four-dimensional non-chiral amplitude can be obtained from six-dimensions by performing a massless reduction, i.e. eq.(\ref{reduction}). In fact the relationship between the two is continuous, due to the fact that the supersymmetry of the six dimensional theory is non-chiral. 

To see how the six-dimensional and the non-chiral four-dimensional amplitude are related, one starts with reduction. Besides (super)momentum delta functions, the six dimensional amplitude are written in terms as rational functions of the following Lorentz invariants: 
\begin{equation}
6D:\quad s_{ij},\;tr\left(\displaystyle{\not}\mathcal{P}_{even}\right),\quad \mathfrak{q}\displaystyle{\not}\mathcal{P}_{even}\tilde{\mathfrak{q}},\quad \mathfrak{q}\displaystyle{\not}\mathcal{P}_{odd}\mathfrak{q},\quad \tilde{\mathfrak{q}}\displaystyle{\not}\mathcal{P}_{odd}\tilde{\mathfrak{q}}\,,
\end{equation}
where $\displaystyle{\not}\mathcal{P}_{even}$ and $\displaystyle{\not}\mathcal{P}_{odd}$ indicates strings of even and odd number of momenta respectively. The purely bosonic invariants reduces trivially to four dimensions, and are all non-vanishing except terms that are proportional to Gram determinants which we will discuss shortly. The supermomenta dependent invariants reduce as 
\begin{eqnarray}
\nonumber&&6D:\;\;\{\mathfrak{q_i}\displaystyle{\not}\mathcal{P}_{even}\tilde{\mathfrak{q}}_j,\;\;\; \mathfrak{q}_i\displaystyle{\not}\mathcal{P}_{odd}\mathfrak{q}_j,\;\;\; \tilde{\mathfrak{q}}_i\displaystyle{\not}\mathcal{P}_{odd}\tilde{\mathfrak{q}}_j\}\\
\nonumber \rightarrow&&4D:\;\;\{q^{1}_i\displaystyle{\not}\mathcal{P}_{even}q^{2}_j+\bar q_{i4}\displaystyle{\not}\mathcal{P}_{even}\bar q_{j3},\;\;\; {q}^1_{i}\displaystyle{\not}\mathcal{P}_{odd}\bar{q}_{j4}+\bar{q}_{i4}\displaystyle{\not}\mathcal{P}_{odd}q^1_j,\;\;\; {q}_{i}^2\displaystyle{\not}\mathcal{P}_{odd}\bar{q}_{j3}+\bar{q}_{i3}\displaystyle{\not}\mathcal{P}_{odd}q^2_j\}\\
\label{4Damp}
\end{eqnarray}
One can see that all possible terms are non-vanishing in four dimensions. 

We now discuss the uplifting. For the four-point, the uplifting is trivial since we have \begin{equation}
4D: \mathcal{A}_4=-i\frac{\delta^4(\sum_i p_i)\delta^4(\sum_j q_j)\delta^4(\sum_l \bar q_l)}{st}\rightarrow6D: \mathcal{A}_4=-i\frac{\delta^6(\sum_i \mathfrak{p}_i)\delta^4(\sum_j \mathfrak{q}_j)\delta^4(\sum_l \bar{\mathfrak{q}}_l)}{st}
\end{equation}
Since using a massive reduction, this gives the massive amplitudes in four dimensions, this implies that the non-chiral amplitude ``knows" about it's massive extension. Beyond four-point, one might expect the uplifting to be more subtle due to the extra supermomenta appearing in the amplitude. However, the subtlety is trivialize once one realizes that the four-dimensional R-symmetry and its non-chirality ensures that all terms combine in such a way that one can trivially combine the four-dimensional supermomenta into six-dimensional ones. The manifest SU(2)$\times$SU(2) R invariance implies that the four-dimensional supermomenta come in anti-symmetrize pairs, $q_j^{[1}q_i^{2]}$ and $\bar{q}_{j[3}\bar{q}_{i4]}$. Furthermore, the non-chiral nature of the amplitude also implies that the amplitude is invariant if one exchanges all $q^1_i\leftrightarrow\bar{q}_{i4}$ and $q^2_i\leftrightarrow\bar{q}_{i3}$. These two constraints plus Lorentz invariance implies that the supermomenta dependent terms come in the combination such that one can directly uplift the expression to six dimensions:
\begin{eqnarray}
\nonumber4D:\;\; q^{[1}_i\displaystyle{\not}\mathcal{P}_{even}q^{2]}_j+\bar q_{i[4}\displaystyle{\not}\mathcal{P}_{even}\bar q_{j3]}\quad\quad\quad\quad&\rightarrow&\; 6D:\;\;\mathfrak{q_i}\displaystyle{\not}\mathcal{P}_{even}\tilde{\mathfrak{q}}_j+\tilde{\mathfrak{q_i}}\displaystyle{\not}\mathcal{P}_{even}\mathfrak{q}_j\\
\nonumber 4D:\;\; ({q}^{[1}_{i}\displaystyle{\not}\mathcal{P}_{odd}\bar{q}_{j[4}+\bar{q}_{i[4}\displaystyle{\not}\mathcal{P}_{odd}q^{[1}_j)({q}_{i}^{2]}\displaystyle{\not}\mathcal{P}_{odd}\bar{q}_{j3]}+\bar{q}_{i3]}\displaystyle{\not}\mathcal{P}_{odd}q^{2]}_j)&\rightarrow&\;6D:\;\;\mathfrak{q}_i\displaystyle{\not}\mathcal{P}_{odd}\mathfrak{q}_j\tilde{\mathfrak{q}}_i\displaystyle{\not}\mathcal{P}_{odd}\tilde{\mathfrak{q}}_j\,\\
\end{eqnarray}
where the uplift to six dimensions is done according to the map eq.(\ref{4Damp}) and eq.(\ref{reduction}). Interestingly, the fact that the four-dimensional amplitude is written in projective superspace translate to chiral symmetry in six dimensions, i.e. symmetry under the exchange of $\mathfrak{q}_i\leftrightarrow \tilde{\mathfrak{q}}_i$, or equivalently, that the amplitude can be written directly in seven dimensions where the two chirality spinors are combined. 

There are potential six-dimensional Lorentz invariant objects that vanish in four dimensions and thus would invalidate the equivalence, i.e. the  Levi-Cevita tensors and Gram determinants. Due to the non-chiral nature of the six-dimensional theory, the on-shell amplitude will always be symmetric under $\mathfrak{q}_i\leftrightarrow \tilde{\mathfrak{q}}_i$ and hence it will not contain the chirality matrix $\gamma_7$ or equivalently, Levi-Cevita tensors. Thus one would only need to worry about Gram determinants. Such objects are not expected to appear at tree level, since any term with a Gram determinant will have to be by itself DSC covariant with the correct inversion weight along with the correct multi-particle poles. Generically this implies that such terms would have double propagators in order for it to achieve the inversion weight dictated in eq.(\ref{invert}), and hence would be ruled out by factorization constraints. However,  at the moment we do not have rigourous proof of their non-existence at tree-level. Thus the precise statement is that the uplifting captures the six dimensional amplitude modulo terms proportional to Gram determinants.

This provides a convenient way of obtaining massive amplitude. One simply start with the four dimensional result and use the above identification to combine the supermomenta into the six dimensional version along with continuing all the vector indices to six dimensions. This gives the six-dimensional amplitude. One then uses the fact that a six-dimensional massless momenta is equivalent to a four dimensional massive one:
 \begin{equation}
  \mathfrak{p}^2 = p^2 - p_4^2-p_5^2 \equiv p^2 - m \tilde m = 0\,,  
\end{equation}
one can obtain massive kinematics in four dimensions by identifying~\cite{Bern:2010qa}:
\begin{eqnarray}
	\lambda^A{}_a=\left(\begin{array}{cc}
			-\kappa\mu_\alpha&\lambda_\alpha\\ 
			\tilde\lambda^{\dot\alpha}&\tilde\kappa\tilde\mu^{\dot\alpha}
		\end{array}\right)\,,\qquad 
&&	
	\tilde \lambda_{A\dot a}=\left(\begin{array}{cc}
		\kappa'\mu^\alpha  &\lambda^\alpha\\ 
		-\tilde\lambda_{\dot\alpha}&\tilde\kappa'\tilde\mu_{\dot\alpha} 
		\end{array}\right)\,,
\end{eqnarray}
where 
\begin{eqnarray}
&&p_{\alpha\dot\alpha} = \lambda_\alpha\tilde\lambda_{\dot\alpha}+\rho\,\mu_\alpha\tilde\mu_{\dot\alpha} \,,
\end{eqnarray}
and 
\begin{eqnarray}
	&&\quad\rho=\kappa\tilde\kappa=\kappa' \tilde\kappa' \,, \hskip .5 cm 
          \kappa\equiv\frac{m}{\langle\lambda\mu\rangle}\,, \hskip .5 cm 
	  \tilde\kappa=\frac{\tilde m}{[\mu\lambda]}\,, \hskip .5 cm 
	  \kappa'\equiv\frac{\tilde m}{\langle\lambda\mu\rangle}\,, \hskip .5 cm 
	  \tilde\kappa'=\frac{ m}{[\mu\lambda]}\,. \hskip .5 cm 
\end{eqnarray}
As an example, the four-point massive amplitude using the above reduction is given as: 
\begin{eqnarray}
\nonumber 4D\; massive: \mathcal{A}_4&=&\frac{-i}{st}\delta^{(2)}\left(\sum^4_{i=1} \lambda_{i\alpha} \eta_{i}^1+\kappa_i \mu_{i\alpha}\bar{\eta}_{i4}\right)\delta^{(2)}\left(\sum^4_ {j=1}\tilde{\kappa}_j\tilde{\mu}_j^{\dot{\alpha}}\eta^1_{j}-\tilde{\lambda}_j^{\dot{\alpha}}\bar\eta_{j4}\right)\\
&&\times\delta^{(2)}\left(\sum_{k=1}^4 \kappa^{'}_{k}\mu_k^{\alpha}\bar{\eta}_{k3}+\lambda^\alpha_k\eta^{2}_k \right)\delta^{(2)}\left(\sum^4_{l=1}-\tilde{\lambda}_{l\dot{\alpha}}\bar\eta_{l3}+\tilde\kappa^{'}_l\tilde{\mu}_{l\dot{\alpha}}\eta^2_{l}\right)\,.
\end{eqnarray}
where $\delta^{(2)}$ indicates that the SL(2,C) index $\alpha,\dot\alpha$ is summed over. The component states are the coefficients in the $\eta,\bar\eta$ expansion of a scalar superfield. This scalar superfield is simply a half-Fourier transform of the original chiral superfield:
\begin{equation}
\Phi^{D=4}_{Projective}(\eta,\bar{\eta})=\int d\eta^{3}d\eta^{4}e^{\eta^3\bar{\eta}_3+\eta^4\bar{\eta}_4}G^{D=4}_{chiral}(\eta)\,.
\end{equation}
The superfield is a scalar since the leading component field in the $(\eta,\bar{\eta})$ expansion is the scalar $\phi^{12}$. Starting with the following chiral expansion 
\begin{equation}
G^{D=4}_{chiral}(\eta)=G^++\eta^I\psi_I+\frac{1}{2}\epsilon_{IJKL}\eta^I\eta^J\phi^{KL}+\frac{1}{3!}\epsilon_{IJKL}\eta^I\eta^J\eta^K\bar\psi^{L}+\frac{1}{4!}\epsilon_{IJKL}\eta^I\eta^J\eta^K\eta^{L}G^-\,,
\end{equation}
we obtain the following component expansion of the scalar superfield 
\begin{eqnarray}
\nonumber\Phi^{D=4}_{Projective}(\eta,\bar{\eta})&=&\phi^{21}+\eta^{m'}\bar\psi_{m'}+\bar\eta_m\psi^m+\bar\eta_{m}\eta^{n'}\phi_{n'}\,^m+\eta^2G^-+\bar\eta^2G^+\\
&&+\eta^2\bar\eta_m\bar\psi^m+\bar\eta^2\eta^{m'}\psi_{m'}+\frac{1}{4}\bar\eta^2\eta^2\phi^{43}
\end{eqnarray}
where now $\eta^2=\frac{1}{2}\eta^{m'}\eta_{m'}, \;\bar\eta^2=\frac{1}{2}\bar\eta_{m}\bar\eta^{m}$, and the raising and lowering of indices is performed through SU(2) metric $\epsilon^{12}=\epsilon_{21}=1$, and we've identified $I=(3,4)\rightarrow m=(1,2)$. The projective superfield can be directly mapped into the six-dimensional superfield~\cite{DHS} 
\begin{eqnarray}
\Phi^{D=6}(\eta,\tilde{\eta}) &=&
  \phi 
  + \chi^a \eta_a 
  + \tilde{\chi}_{\da}\tilde{\eta}^{\da} 
   + \phi'(\eta)^2 
  + g^a\,_{\da}\eta_a\tilde{\eta}^{\da}
  + \phi''(\tilde{\eta})^2
  \nn\\
  && \null
   + \tilde{\lambda}_{\da}(\eta)^2\tilde{\eta}^{\da}
  + \lambda^a\eta_a(\tilde{\eta})^2
  + \phi'''(\eta)^2(\tilde{\eta})^2 \,,
\end{eqnarray}
where $\eta^2=\frac{1}{2}\eta^a\eta_a,\;\;\tilde\eta^2=\frac{1}{2}\tilde\eta_{\dot a}\tilde\eta^{\dot a}$. This gives the following map between four-dimensional states to six-dimensional ones:
\begin{eqnarray}
\nonumber 6D:\quad\left(\phi,\;\phi',\;\phi'',\;\phi'''\right)&\leftrightarrow&\;4D:\quad\left(\phi^{21},\;\phi^{42},\;\phi^{13},\;\phi^{34}\right)\\
\nonumber 6D:\quad \left(\begin{array}{cc} g^1\,_{\dot 1} & g^1\,_{\dot 2} \\ g^2\,_{\dot 1} &  g^2\,_{\dot 2}\end{array}\right)\quad&\leftrightarrow&\;4D:\quad\left(\begin{array}{cc}\phi^{23} & -G^-  \\  -G^+ & \phi^{14}\end{array}\right)\\
\nonumber 6D:\quad \chi^a,\;\tilde\chi_{\dot a},\;\lambda^a,\;\tilde\lambda_{\dot a}\quad&\leftrightarrow&\;4D:\quad\left(\begin{array}{c}\bar\psi^2 \\ \psi_3\end{array}\right),\;\left(\begin{array}{c} -\psi_4 \\ -\bar\psi^1\end{array}\right),\;\left(\begin{array}{c}-\bar\psi^3 \\ \psi_2\end{array}\right),\,\left(\begin{array}{c}-\psi_1 \\ \bar\psi^4\end{array}\right)\,.\\
\end{eqnarray}

\subsection{Chiral $\rightarrow$ non-chiral amplitudes}
In this subsection we will derive the non-chiral amplitudes directly from the chiral ones by performing half-Fourier transforms. The half-Fourier transformed result can be easily obtained by noting that for general $n$-point amplitude, the chiral representation can be obtained from the anti-chiral representation by Fourier transforming the $\bar\eta$s to $\eta$s. In particular, one has 
\begin{eqnarray}
\nonumber\mathcal{A}_n(N^{(n-4)}MHV)&=&\left[\int \prod_{i=1}^nd\bar\eta_i^Ie^{\eta^I\bar\eta_I}\right]^4\mathcal{A}_n(\overline{\rm MHV})\\
&=&\left[\int \prod_{i=1}^nd\bar\eta_i^Ie^{\eta^I\bar\eta_I}\delta^2(\sum_{i=1}^{n}\bar{q}_{iI\dot a})\right]^4\frac{1}{\prod_{i=1}^n[ii+1]}
\end{eqnarray}
where the repeated $I$ are not summed over and the fourth power indicates the Fourier transform for each of the four $I$ R-index. From the above equation one can deduce that $\mathcal{A}_n(N^{(n-4)}MHV)$ can be written as 
\begin{eqnarray}
\mathcal{A}_n(N^{(n-4)}MHV)=\frac{\left[\delta^2(\sum_{i=1}^{n}q_{i}^{I\alpha})g(\lambda, x, \theta)\right]^4}{\prod_{i=1}^n[ii+1]}\,.
\end{eqnarray}
where $g(\lambda, x, \theta)$ is some dual conformal invariant function dependent only on the variables $(\lambda, x, \theta)$. This then implies that a single Fourier transformation is simply given by 
\begin{eqnarray}
\delta^2(\sum_{i=1}^{n}q_{i}^{I\alpha})g(\lambda, x, \theta)=\int \prod_{i=1}^nd\bar\eta_i^Ie^{\eta^I\bar\eta_I}\delta^2(\sum_{i=1}^{n}\bar{q}_{iI\dot a})\,.
\end{eqnarray}
The non-chiral form of the $\overline{\rm MHV}$ amplitude is given by Fourier transformation of $\bar\eta_1,\bar\eta_2$ variables. Using the above single Fourier-transformed result one obtains:
\begin{eqnarray}
\mathcal{A}_n(\overline{\rm MHV})=\frac{\delta^4(\sum_{i=1}^{n}q_{i}^{m'\alpha})\delta^4(\sum_{i=1}^{n}\bar{q}_{im\dot a})\left[g(\lambda, x, \theta)\right]^2}{\prod_{i=1}^n[ii+1]}\,.
\end{eqnarray}
Similar results can be derived for MHV amplitudes. The remaining step is to convert $\left[g(\lambda, x, \theta)\right]^2/\prod_{i=1}^n[ii+1]$ into functions of the projective coordinates $(x,\theta,\bar\theta)$ which we will discuss in detail, beginning with the five-point amplitude. 

\subsubsection{Five-point amplitude in non-chiral superspace\label{5pt}}

We will now deduce the form of $f^{D=4}_5$ using half-Fourier transforms. We first note that since the NMHV five-point amplitude in the chiral representation is equivalent to the Fourier transformation of the $\overline{\rm MHV}$ amplitude in the anti-chiral representation, one has\begin{eqnarray}
\frac{\left[\delta^2(\sum_i \lambda_{i\mu}\eta^I_{i})\right]^4}{\langle12\rangle\langle23\rangle\langle34\rangle\langle45\rangle\langle51\rangle}R_{5,42}=\frac{\left[\int \left(\prod^5_{i=1}d\bar{\eta}_{iJ}\right)e^{\eta_i^J\bar\eta_{iJ}}\delta^2(\sum_i \tilde\lambda_{i\dot\mu}\bar\eta_{iI})\right]^4}{[12][23][34][45][51]}
\label{identity}
\end{eqnarray}
where the DSC invariant function $R_{5,42}$ is given in~\cite{DualConformal2}, and the fourth-power of the numerator reflects four products of the same object with different R-index I. Generically $R_{r,st}$ contains spurious poles, however at five point it can be rewritten in a form that has only two particle poles:
\begin{equation}
R_{5,42}=\frac{\langle12\rangle\langle23\rangle\langle34\rangle\langle45\rangle\langle51\rangle\left(\Xi^I_{5,24}\right)^4}{[12][23][34][45][51](\langle45\rangle\langle23\rangle\langle51\rangle)^4}\,.
\end{equation}
where $\Xi^I_{5,24}=\langle 5|\left(x_{52}x_{24}\theta_4^I+x_{54}x_{42}\theta_2^I+x^2_{42}\theta_5^I\right)$. Using the above representation of $R_{5,42}$, eq.(\ref{identity}) can be rewritten as a product of four independent Fourier transformations. This then leads to the following identity 
\begin{equation}
\left[\int \left(\prod^5_{i=1}d\bar{\eta}_{iI}\right)e^{\eta_i^I\bar\eta_{iI}}\delta^2(\sum_i \tilde\lambda_{i\dot\mu}\bar\eta_{iI})\right]=\left[\frac{\delta^2(\sum_i \lambda_{i\mu}\eta^I_{i})}{\langle 45\rangle\langle 23\rangle\langle 51\rangle}\Xi^I_{5,24}\right]\,.
\label{1stidentity}
\end{equation}
As promised, the function $\Xi^I_{5,24}/\langle 45\rangle\langle 23\rangle\langle 51\rangle$ is a dual conformal invariant if one notes that the spinors has the following inversion rules:
\begin{equation}
I[\lambda_{i}^{\alpha}]=\frac{x_i^{\dot\alpha \beta}\lambda_{i\beta}}{\sqrt{x_i^2x^2_{i+1}}},\;\;I[\tilde\lambda_{i\dot\alpha}]=\frac{x_{\alpha\dot\beta}\tilde\lambda_{i}^{\dot\beta}}{\sqrt{x_i^2x^2_{i+1}}}\,.
\end{equation} 
The above inversion rules are symmetric between the chiral and anti-chiral spinors. 

The identity in eq.(\ref{1stidentity}) allows us to straight forwardly extract the half Fourier transformed result:
\begin{eqnarray}
 f^{D=4}_5(\overline{MHV})=\frac{1}{s_{23}s_{45}s_{51}}\left[\frac{\Xi^{m'}_{5,24}\Xi_{5,24m'}}{\langle 45\rangle\langle 23\rangle\langle 51\rangle[12][34]}\right].
 \label{5ptMHV}
\end{eqnarray} 
We would now like to convert the above function into purely projective superspace coordinates. To do so note that one would have the remove the spinor brackets in the denominator in such a way that does not introduce spinor traces, since such objects do not lift to six dimensions. Thus one simply multiply and divide the above result by the (anti-)holomorphic partner of each the spinor brackets. This leads to; 
\begin{eqnarray}
\nonumber f^{D=4}_5&=&\frac{1}{x^2_{13}x^2_{24}x^2_{35}x^2_{41}x^2_{52}}\left[\frac{\Xi^{m'}_{5,24}[5|x_{12}x_{23}x_{34}x_{41}|5]\Xi_{5,24m'}+\bar{\Xi}_{5,24m}\langle 5|x_{12}x_{23}x_{34}x_{41}|5\rangle \bar{\Xi}_{5,24}^m}{x^2_{52}x^2_{41}x^2_{24}}\right].\\
&=&\frac{1}{x^2_{13}x^2_{24}x^2_{35}x^2_{41}x^2_{52}}\left[\frac{\Theta_{5,24}x_{51}x_{12}x_{23}x_{34}x_{41}x_{15}\Theta_{5,24}}{x^2_{52}x^2_{41}x^2_{24}}\right].
\label{f5}
\end{eqnarray} 
where we've added the half-Fourier transform of the MHV amplitude as well to give the full five-point amplitude. We've defined 
\begin{equation}
\Theta_{5,24}=\left[x_{52}x_{24}(\theta_{m'}+\bar\theta_{m})_{45}+x_{54}x_{42}(\theta_{m'}+\bar\theta_{m})_{25}\right]\,,
\end{equation}
and the two $\Theta$ across the gamma trace have their SU(2)-R indices contracted as $^m\,_m$ and $_{m'}\,^{m'}$. One can check that $ f^{D=4}_5$ indeed inverts as eq.(\ref{invert}). 

We are now ready to lift $f_5^{D=4}$ to  $f_5^{D=6}$. The bosonic invariants $x^2_{ij}$ extends trivially to six-dimensions, thus the only non-trivial piece in the uplifting is the supermomenta dependent terms. One only needs to note that under the reduction formula eq.(\ref{reduction}), substituting $(\mathfrak{p},\mathfrak{q},\tilde{\mathfrak{q}})$ with $(x,\theta^{(6)},\tilde\theta^{(6)})$ one has 
\begin{equation}
\bar\theta_i^m\displaystyle{\not}x_{even}\bar\theta_{jm}+\theta_{im'}\displaystyle{\not}x_{even}\theta_{j}^{m'}=\theta_i^{(6)}\displaystyle{\not}x_{even}\tilde{\theta}_j^{(6)}+\tilde{\theta}_i^{(6)}\displaystyle{\not}x_{even}\theta_j^{(6)}
\end{equation}
where ($\theta^{(6)}, \tilde{\theta}^{(6)}$) are the six dimensional dual variables. Looking at eq.(\ref{f5}) one sees that the four-dimensional non-chiral result combines neatly into the form given above, and hence one can directly uplift the result to six-dimensions where one obtains: 
\begin{eqnarray}
\nonumber f_5^{D=6}&=&\frac{\left[(\theta^{(6)}_{54}x_{42}x_{25}+\theta^{(6)}_{52}x_{24}x_{45})x_{51}x_{12}x_{23}x_{34}x_{41}x_{15}(x_{52}x_{24}\tilde\theta_{45}+x_{54}x_{42}\tilde\theta_{25})\right]}{x^2_{13}x^2_{24}x^2_{35}x^2_{41}x^2_{52}x^2_{52}x^2_{41}x^2_{24}}\\
&&+(\theta\leftrightarrow\tilde\theta).
\end{eqnarray} 
Translating back to the six dimensional on-shell space using eq.(\ref{sixd}), one indeed obtains the five-point amplitude given in~\cite{DHS}.

Note that in the non-chiral representation, one can still differentiate between different N$^K$MHV terms. From analyzing the half Fourier transformation, one can conclude that a N$^K$MHV amplitude has the fermionic dependence in the form of 
\begin{equation}
f_n^{D=4}|_{N^KMHV}\sim \theta^{2(K)}\bar{\theta}^{2(n-4-K)}
\end{equation}
where $|_{N^KMHV}$ means we're picking out the N$^K$MHV piece. Indeed one sees that the MHV and $\overline{\rm MHV}$ amplitude corresponds to $\theta x\cdot\cdot x\theta$ and $\bar\theta x\cdot\cdot x\bar\theta$ terms in $ f^{D=4}_5$ respectively.\footnote{There are no $\theta x\cdot\cdot x\bar\theta$ terms since there are even number of $x$s in between.} Note that this does not contradict the statement that the six dimensional parent amplitude does not differentiate between different helicities, since an chiral and anti-chiral spinor in four dimensions combine into a singe six dimensional spinor as shown in eq.(\ref{reduction}).

\subsubsection{six-point $\overline{\rm MHV}$ amplitude in non-chiral superspace.}
As a further exercise, lets consider the six-point $\overline{\rm MHV}$ amplitude in the non-chiral representation. This amplitude corresponds to the N$^2$MHV in the chiral representation and is given as~\cite{Drummond:2008cr}:
\begin{eqnarray}
A^{\overline{\rm MHV}}_6=\frac{\delta^4(P)\delta^8(Q)}{\prod_{i=1}^6\langle ii+1\rangle}R_{6;25}R^{0;25}_{6;52;35}\,,
\label{MHVbar}
\end{eqnarray}
where the generalized R-function is defined in~\cite{Drummond:2008cr}, and we give for our case:
\begin{eqnarray}
R^{0;25}_{6;52;35}=\frac{\langle 32\rangle\langle\hat{5}4\rangle\left[\langle\xi|x_{23}x_{35}\theta_{52}+\langle\xi|x_{25}x_{53}\theta_{32}\right]^4}{x^2_{35}\langle \xi|x_{23}x_{35}|\hat{5}\rangle\langle \xi|x_{23}x_{35}|4\rangle\langle \xi|x_{25}x_{53}|3\rangle\langle \xi|x_{25}x_{53}|2\rangle}\,,
\end{eqnarray}
whith $\langle\xi|=\langle6|x_{65}x_{52}$ and $\langle\hat5|=\langle6|x_{62}x_{25}$. As with the five-point case, the (super) momentum delta function independent part in eq.(\ref{MHVbar}) is secretly a fourth power of a single function divided by the anti-holomorphic spinor string $\prod_{i=1}^6[ ii+1]$. More explicitly, one can show that 
\begin{eqnarray}
\frac{R_{6;25}R^{0;25}_{6;52;35}}{\prod_{i=1}^6\langle ii+1\rangle}=\frac{1}{\prod_{i=1}^6[ ii+1]}\left[\frac{\Xi_{6;25}\Xi_{\xi2;35}}{\langle 34\rangle\langle61\rangle\langle56\rangle\langle6|x_{65}x_{52}|2\rangle x^2_{25}}\right]^4
\end{eqnarray}  
where we use $\Xi_{\xi2;35}$ to denote $\langle\xi|x_{23}x_{35}\theta_{52}+\langle\xi|x_{25}x_{53}\theta_{32}$. Following the same arguments as presented in the previous subsection, one can use the above formula to derive the half-Fourier transformed result of the $\overline{\rm MHV}$ amplitude in the anti-chiral basis to obtain the non-chiral form. One then finds:
\begin{eqnarray}
\nonumber A^{\overline{\rm MHV}}_6&=&\frac{\delta^4(x_{1n+1})\delta^4(\theta_{1n+1})\delta^4(\bar{\theta}_{1n+1})}{\prod_{i=1}^6[ ii+1]}\left[\frac{\Xi_{6;25}\Xi_{\xi2;35}}{\langle 34\rangle\langle61\rangle\langle56\rangle\langle6|x_{65}x_{52}|2\rangle x^2_{25}}\right]^2\\
\nonumber&=&\frac{\delta^4(x_{1n+1})\delta^4(\theta_{1n+1})\delta^4(\bar{\theta}_{1n+1})}{\prod_{i=1}^6s_{ii+1}}\frac{[34][61][56]\langle 54\rangle\langle23\rangle\langle21\rangle(\Xi_{6;25})^2(\Xi_{\xi2;35})^2}{x^2_{35}x^2_{62}x^2_{51}\langle6|x_{65}x_{52}|2\rangle^2(x^2_{25})^2}
\label{6ptMHV}
\end{eqnarray}  
Again, we would like to write the above in purely projective superspace coordinates, thus we multiply and divide the above function by the anti-holomorphic partner of the spinor string in the denominator, i.e. $[6|x_{65}x_{52}|2]$. After some simple algebra, we arrive at 
\begin{eqnarray}
\nonumber A^{\overline{\rm MHV}}_6&=&\frac{\delta^4(x_{1n+1})\delta^4(\theta_{1n+1})\delta^4(\bar{\theta}_{1n+1})}{\prod_{i=1}^6s_{ii+1}}\frac{1}{x^2_{35}x^2_{62}x^2_{51}tr(x_{32}x_{26}x_{65}x_{53})^2(x^2_{25})^2}\\
&&\left(\Theta^{m'}_{6;25}x_{61}x_{12}x_{23}x_{35}x_{51}x_{16}\Theta_{m'6;25}\Theta^{n'}_{6\xi;35}x_{64}x_{43}x_{32}x_{26}\Theta_{n'6\xi;35}\right)
\end{eqnarray}
where $\Theta_{m'6;25}$ is defined in the previous subsection and $\Theta^{n'}_{6\xi;35}=x_{65}x_{52}(x_{23}x_{35}\theta_{52}+x_{25}x_{53}\theta_{32})$. For the MHV amplitude one simply trade $\theta$ with $\bar{\theta}$. 

\subsubsection{The $n$-point ${\rm MHV}-\overline{\rm MHV}$ amplitude in non-chiral superspace.}
From the discussion so far, one sees that for the MHV and $\overline{\rm MHV}$ amplitude, the half-Fourier transformed result can be obtained quite easily and one is only a step away from the fully non-chiral representation. The remaining step is simply multiply and divide by the (anti-) holomorphic partner of what ever spinor string appears in the denominator. Thus for the form of the amplitude in projective superspace, it is sufficient to give the half-Fourier transformed result prior to removing the spinor inner products. 

The general $n$-point formula for the half-Fourier transformed result can be written as:
\begin{eqnarray}
\nonumber && f_n^{D=4}(\overline{\rm MHV})=\frac{1}{\prod_{i=1}^n[ ii+1]}\left[ \frac{\left(\langle n|x_{n(n-1)}x_{(n-1)2}\theta_{2n}+\langle n|x_{n2}x_{2(n-1)}\theta_{(n-1)n}\right)^2}{(\langle n1\rangle\langle nn-1\rangle\langle n-2n-3\rangle)^2}\right.   \\
\nonumber&&\left.\times\prod_{r=1}^{n-5}\frac{\left(\langle\xi_r|x_{(r+1)(n-5+r)}x_{(n-5+r)(n-1)}\theta_{(n-1)(r+1)}+\langle\xi_r|x_{(r+1)(n-1)}x_{(n-1)(n-5+r)}\theta_{(n-5+r)(r+1)}\right)^2}{\left(\langle\xi_r(1+r)\rangle x^2_{(n-1)(r+1)}\right)^2} \right]\,.
\end{eqnarray} 
The notations are based on Henn and Drummonds all tree solution~\cite{Drummond:2008cr}. The spinor $\langle \xi_r|$ is given by 
\begin{equation}
\langle\xi_r|=\langle n|x_{n(n-1)}x_{(n-1)(r+1)}
\end{equation}
One can see that the above formula reproduces eq.(\ref{5ptMHV}) and eq.(\ref{6ptMHV}). As a further demonstration we give the seven-point result:
\begin{eqnarray}
\nonumber && f_7^{D=4}(\overline{\rm MHV})=\frac{\left(\Xi_{7;26}\right)^2}{\prod_{i=1}^7[ ii+1]}\left[ \frac{\left(\langle\xi_1|x_{23}x_{36}\theta_{62}+\langle\xi_1|x_{26}x_{63}\theta_{32}\right)^2\left(\langle\xi_2|x_{34}x_{46}\theta_{63}+\langle\xi_2|x_{36}x_{64}\theta_{43}\right)^2}{(\langle 71\rangle\langle 76\rangle\langle 54\rangle \langle7|x_{76}x_{63}|3\rangle\langle7|x_{76}x_{62}|2\rangle x^2_{62}x^2_{63})^2}\right] \,.  \\
\end{eqnarray} 
Once converted into the full superspace, one can continue the result to six dimensions, where through massive decomposition one can obtain the four dimensional massive amplitude. 

\subsection{Non-chiral amplitudes in chiral momentum twistor space}
The fact that $f_5^{D=4}$ in eq.(\ref{5ptMHV}) is not manifest cyclic invariant, is not surprising since full cyclic symmetry of the R-function in the chiral representation is only manifest using momentum twistors. Another related issue is that one does not expect that higher point amplitudes to be products of generalizations of $f^{D=4}_5$ in the form given in eq.(\ref{5ptMHV}) due to the lack of spurious poles. As discussed earlier, the original form of $R_{r,st}$ contains spurious poles such as $1/\langle r|x_{rs}x_{st}|t\rangle$ which degenerates for $R_{5,24}$ to $1/\langle 51\rangle[51]\langle54\rangle$. In this subsection, we will rewrite $f_5^{D=4}$ in momentum twistor space which realizes the hidden cyclic symmetry and provides a form where the generalization to higher points introduces spurious poles.

Due to the non-chiral nature of the dual space time, it will be natural for us to introduce momentum twistors of both chiralities. Furthermore since the dual symmetry is non-chiral, this will have subtle consequences in the proper definition of dual conformal invariants. Recall that the usual momentum twistor $\mathcal{Z}_i^A=\left(\lambda_i,\;\mu_i,\;\chi_i\right)$ is used to define dual momentum chiral superspace via the incident relation $\mu_i^{\dot\alpha}=x_i^{\dot\alpha\alpha}\lambda_{i\alpha}=x_{i+1}^{\dot\alpha\alpha}\lambda_{i\alpha}$ and $\chi_{iI}=\theta^{\alpha}_{iI}\lambda_{i\alpha}=\theta^{\alpha}_{i+1I}\lambda_{i\alpha}$. For the non-chiral superspace, we instead use both $\mathcal{Z}_i^A=\left(\lambda_i,\;\mu_i,\;\chi_{im'}\right)$ and $\mathcal{W}_{iA}=(\tilde\mu_i,\;\tilde\lambda_i,\;\bar\chi^m_{i})$ along with a new set of incident relations
\begin{eqnarray}
\nonumber&&\tilde\mu_i^{\alpha}=x_i^{\dot\alpha\alpha}\tilde\lambda_{i\dot\alpha}=x_{i+1}^{\dot\alpha\alpha}\tilde\lambda_{i\dot\alpha}\\
&&\bar\chi^m_{i}=\bar{\theta}^{m\dot\alpha}_i\tilde\lambda_{i\dot\alpha}=\bar\theta^{m\dot\alpha}_{i+1}\tilde\lambda_{i\dot\alpha}\,.
\end{eqnarray}
In this ``ambi"-momentum space formalism, there are three types of dual conformal covariants:
\begin{equation}
\langle ijkl\rangle=\epsilon_{ABCD}Z^A_iZ^B_jZ^C_kZ^D_l, \quad\left[ijkl\right]=\epsilon^{ABCD}W_{Ai}W_{jB}W_{kC}W_{lD},\quad\left( ij\right)=Z^A_iW_{jA}\,.
\label{invariants} 
\end{equation}
where $Z$ and $W$ are just the bosonic part of the supertwistors. The second type is simply the anti-chiral partner of the first type,\footnote{That is, we have for the special case $\left[i-1 ij-1j\right]=[i-1i][j-1j]x^2_{ij}$.} while the third type is equivalent to
\begin{equation}
\left( ij\right)=\lambda^\alpha_{i}\tilde\mu_{j\alpha}+\mu_i^{\dot\alpha}\tilde\lambda_{j\dot\alpha}=\langle i|x_{ij}|j]=\frac{\langle ij-1jj+1\rangle}{\langle j-1j\rangle\langle jj+1\rangle}\,.
\end{equation}
From this we can deduce 
\begin{equation}
W_{jA}=\frac{\epsilon_{ABCD}Z^B_{j-1}Z^C_jZ^D_{j+1}}{\langle j-1j\rangle\langle jj+1\rangle}\,.
\end{equation}

Note that we refer to the objects in eq.(\ref{invariants}) as dual conformal ``covariants". This differs from the usual chiral DSC symmetry, where these objects will be considered as invariants. The discrepancy lies in the fact that the non-chiral conformal boost generator given in eq.(\ref{extension}) combined with $K=IPI$, implies that  the spinors invert under non-chiral dual conformal inversion as 
\begin{equation}
I^{nc}[\lambda_{i}^{\alpha}]=\frac{x_i^{\dot\alpha \beta}\lambda_{i\beta}}{\sqrt{x_i^2x^2_{i+1}}},\;\;I^{nc}[\tilde\lambda_{i\dot\alpha}]=\frac{x_{\alpha\dot\beta}\tilde\lambda_{i}^{\dot\beta}}{\sqrt{x_i^2x^2_{i+1}}}\,,
\end{equation} 
where we use the superscript $c$ and $nc$ to denote chiral and non-chiral respectively. This differs from the chiral inversion rules:
\begin{equation}
I^c[\lambda_{i}^{\alpha}]=x_i^{\dot\alpha \beta}\lambda_{i\beta},\;\;I^c[\tilde\lambda_{i\dot\alpha}]=x_{\alpha\dot\beta}\tilde\lambda_{i}^{\dot\beta}\,.
\end{equation} 
The consequence is that while the chiral inversion simply exchanges $\mu$ and $\lambda$, the non-chiral inversion gives
\begin{equation}
I^{nc}[\lambda_{i}^{\alpha}]=\frac{\mu_i^{\dot\alpha}}{\sqrt{x_i^2x^2_{i+1}}},\;\;I^{nc}[\mu_{i}^{\dot\alpha}]=\frac{\lambda_i^{\alpha}}{\sqrt{x_i^2x^2_{i+1}}}\,.
\end{equation}
Thus the objects in eq.(\ref{invariants}) are covariant under this discrete symmetry:
\begin{equation}
I^{nc}[\langle ijkl\rangle]=\frac{\langle ijkl\rangle}{\sqrt{x_i^2x^2_{i+1}x_j^2x^2_{j+1}x_k^2x^2_{k+1}x_l^2x^2_{l+1}}},\quad I^{nc}[\left( ij\right)]=\frac{\left( ij\right)}{\sqrt{x_i^2x^2_{i+1}x_j^2x^2_{j+1}}}\,.
\end{equation}
and similar results for $\left[ijkl\right]$. Thus non-chiral dual conformal invariants can be constructed using these covariant objects with the extra constraint that the inversion weights cancel properly. 

We are now ready to rewrite $f_5^{D=4}$ in momentum twistor space. Using the fact that~\cite{Hodges:2009hk, Mason:2009qx, ArkaniHamed:2009vw} 
\begin{equation}
\Xi^I_{r,st}=\frac{\left(\langle r-1,r,s-1,s\rangle\chi^I_t+{\rm cyclic}\right)}{\langle s s-1\rangle\langle t t-1\rangle}
\end{equation} 
we see that eq.(\ref{f5}) can be rewritten as 
\begin{eqnarray}
\nonumber f_{5}^{D=4}&=&\frac{1}{\prod_{i=1}^5[ii+1]}\left[\frac{\left(\langle 1,2,3,4\rangle\chi_5+{\rm cyclic}\right)^{m'}\left(\langle 1,2,3,4\rangle\chi_5+{\rm cyclic}\right)_{m'}}{(\langle 12\rangle\langle 23\rangle\langle 34\rangle\langle 45\rangle\langle 51\rangle)^2}\right]\\
\nonumber&&+\frac{1}{\prod_{i=1}^5\langle ii+1\rangle}\left[\frac{\left([1,2,3,4]\bar\chi_5+{\rm cyclic}\right)_m\left([1,2,3,4]\bar\chi_5+{\rm cyclic}\right)^m}{([12][23][34][45][51])^2}\right]\\
\nonumber &=&\frac{1}{s_{12}s_{23}s_{34}s_{45}s_{51}}\left[\frac{\left(\langle 1,2,3,4\rangle\chi_5+{\rm cyclic}\right)^{m'}\left(\langle 1,2,3,4\rangle\chi_5+{\rm cyclic}\right)_{m'}}{\langle 12\rangle\langle 23\rangle\langle 34\rangle\langle 45\rangle\langle 51\rangle}\right.\\
&&\left.+\frac{\left([1,2,3,4]\bar\chi_5+{\rm cyclic}\right)_m\left([1,2,3,4]\bar\chi_5+{\rm cyclic}\right)^m}{[12][23][34][45][51]}\right]
\label{spinor}
\end{eqnarray}
Thus one sees that using momentum twistors, $f_{5}^{D=4}$ is indeed manifestly cyclic symmetric. Note that the terms in the square bracket of the first line is non-chiral DSC invariant, where the spinor strings exactly cancel the nontrivial inversion weight arising from the numerator. 

For higher point amplitudes we define the following non-chiral DSC covariants   
\begin{eqnarray}
\nonumber \mathcal{R}_{n,n-1,2}&=&\frac{\left(\langle n-1,n-2,2,1\rangle\chi_n+{\rm cyclic}\right)}{\langle21\rangle\langle 1n\rangle\langle nn-1\rangle\langle n-1n-2\rangle\langle n-2n-3\rangle},\\
\nonumber \mathcal{R'}_{U_r,a,b}\;\;&=&\frac{\left(\langle a,a-1,b,b-1\rangle\chi_{U_r}+{\rm cyclic}\right)}{\langle \xi_r r+1\rangle \langle aa-1\rangle\langle bb-1\rangle x^2_{(n-1)(r+1) }}\\
\nonumber\mathcal{\bar{R}}_{n,n-1,2}&=&\frac{\left([  n-1,n-2,2,1]\bar\chi_n+{\rm cyclic}\right)}{[21][1n][nn-1][n-1n-2][ n-2n-3]}\,.\\
\nonumber\mathcal{\bar{R}'}_{\bar{U}_r,a,b}\;\;&=&\frac{\left([  a,a-1,b,b-1]\bar\chi_{\bar{U}_r}+{\rm cyclic}\right)}{[ \xi_r r+1][aa-1][bb-1] x^2_{(n-1)(r+1) }}\,.\\
\end{eqnarray}
where $U_r$ is the twistor defined such that $\langle \xi_r|=U_r^{A}I_{AB}$ and $I_{AB}$ is the infinity twistor. It's explicit form, in terms of $\langle \xi_r|$ can be found in appendix A of ref.~\cite{Mason:2009qx}. Similar definition applies to $\bar{U}_r$. Using these covariants, the MHV and $\overline{\rm MHV}$ amplitude is given as 
\begin{eqnarray}
\nonumber f^{D=4}_{n}(\overline{\rm MHV})&=&\frac{1}{\prod_{i=1}^n[ii+1]} \left[\mathcal{R}_{n,n-1,2}\prod^{n-5}_{r=1} \mathcal{R'}_{U_r;n-1,n-5+r}\right]^2,\\
\nonumber f^{D=4}_{n}({\rm MHV})&=&\frac{1}{\prod_{i=1}^n\langle ii+1\rangle } \left[\mathcal{\bar{R}}_{n,n-1,2}\prod^{n-5}_{r=1} \mathcal{\bar{R}'}_{U_r;n-1,n-5+r}\right]^2
\end{eqnarray}
where the square indicates the contraction of the two SU(2)-R indices. Note that $\mathcal{R}_{r,st}$ is related to $\mathcal{\bar{R}}_{r,st}$ via exchanging $\mathcal{Z}\rightarrow\mathcal{W}$. Note that the combination given in the square bracket are dual conformal invariants. We expect that higher point ${\rm N^KMHV}$ amplitudes can be expressed in terms of products of these $\mathcal{R}_{r,st}$ and $\mathcal{\bar{R}}_{r,st}$, in a fashion where $\mathcal{Z}\leftrightarrow\mathcal{W}$ symmetry is manifest.

\subsection{Toward full DCS representation}
A glaring omission in the discussion of this section so far is the dual coordinates $y_{m'}\,^n$. While it is part of the full dual projective superspace, its presence is nowhere to be found in the explicit form of the amplitude, as well as in the incident relations for the momentum twistors.\footnote{As in three dimensions~\cite{Arthur}, one proposal is to extend the incident relation to 
\begin{eqnarray}
\nonumber\mu^{\dot\alpha}&=&-ix^{\alpha\dot\alpha}\lambda_{\alpha},\;\;\tilde\mu^{\alpha}=-ix^{\alpha\dot\alpha}\tilde\lambda_{\dot\alpha}\\
\bar\chi^m&=&\bar{\theta}^{m\dot\alpha}\tilde\lambda_{\dot\alpha},\;\;
\chi^{m'}=\theta^{m'\alpha}\lambda_{\alpha}\,,\\
\bar\chi_{m'}&=&y_{m'}\,^m\bar\eta_m,\;\;
\chi^{m}=y_{m'}\,^m\eta^{m'}\,
\end{eqnarray} }
The lack of its presence can be traced back to the fact that $f_n^{D=4}$ is not fully DSC covariant. In fact, one can check that by dropping all the $y$ dependence in the generators of the DSC group, the generators still faithfully represent the subgroup that is respected by $f_n^{D=4}$, i.e. eq.(\ref{unbrokenG})\,. This also explains the missing origin of the $y$ coordinates in the six-dimensional parent amplitude, where $f_n^{D=6}$ respects only a subgroup of the six-dimensional DSC symmetry as well, as indicated in eq.(\ref{6Dextension}).

The fact that the non-chiral representation is not written in a form where each term is fully DSC covariant is rather unsatisfactory. In the chiral representation, one way to understand MHV amplitudes is that it is proportional to delta functions in the dual chiral superspace. One may expect that for the non-chiral representation, one should instead separate a term proportional to 
\begin{equation}
\delta^4(x_1-x_{n+1})\delta^4(\theta_1-\theta_{n+1})\delta^4(\bar{\theta}_1-\bar{\theta}_{n+1})\delta^4(y_1-y_{n+1})\,.
\end{equation}  
However, the $y$ coordinates of the amplitude are not really independent coordinates and are expressed in terms of $x,\theta,\bar{\theta}$. To see this, note that from the first three hyperplane constraints in eq.(\ref{hyperplane}), one can deduce 
\begin{equation}
y_{i,i+1}=\bar{\eta}_i\eta_i=\frac{\bar\theta_{ii+1}x_{i+1i+2}\theta_{ii+1}}{x^2_{ii+2}}
\label{support}
\end{equation}
Since the amplitude is DSC invariant, any dependence on $y$ must be in terms of differences in order to be translational invariant under $T_{m}\,^{n'}$. One can then conclude that the dependence on $y$ is completely determined by $x,\theta,\bar{\theta}$ coordinates. This explains the lack of $y$ coordinates in the expression $f^{D=4}_5$. On the other hand, the amplitude can be defined in the full dual superspace keeping in mind that it is non-vanishing only on the support of eq.(\ref{support}), i.e. 
\begin{eqnarray}
\mathcal{A}_n&=&\delta^4(x_1-x_{n+1})\delta^4(\theta_1-\theta_{n+1})\delta^4(\bar{\theta}_1-\bar{\theta}_{n+1})\delta^4(y_1-y_{n+1})\\
&&\times\prod_{i=1}^{n}\left[\delta^4(y_{i,i+1}-\frac{\bar\theta_{ii+1}x_{ii+1}\theta_{ii+1}}{x^2_{ii+2}})\right]f_n
\end{eqnarray}
where $\delta^4(y_1-y_{n+1})$ is now needed to impose cyclic identification. However, this trivial rewriting does not cure the fact that $f_n$ is not fully DSC covariant.

\section{Discussion and conclusion\label{section5}}
In this paper we've analysed the non-chiral formulation of both the dual superconformal (DSC) symmetry as well as the amplitude itself. The non-chiral dual symmetry is most naturally defined in dual projective superspace, which can be converted to the usual on-shell variables via hyperplane constraints in eq.(\ref{hyperplane}). Relating the dual space to the on-shell space, one sees that among the original DSC generators, 16 becomes trivial, 32 corresponds to the original superconformal generators, while 16 becomes Yangian level-1 generators. The remaining can be obtain by combing them with the original superconformal generators.

Identification of the dual space allows to write down the amplitude in a non-chiral fashion. Due to the fact that the $MHV$ sector at a given point is as complicated as the other helicity sectors in the non-chiral representation, it is no longer reasonable to separate the amplitude in the form of 
\begin{equation}
\mathcal{A}_n^{N^kMHV}= \mathcal{A}_n^{MHV}\times \mathcal{P}_n^{N^kMHV}\,.
\end{equation} 
Instead, the simplest choice is to separate the momentum and supermomentum delta functions: 
\begin{equation}
\mathcal{A}_n=\delta^4(x_1-x_{n+1})\delta^4(\theta_1-\theta_{n+1})\delta^4(\bar{\theta}_1-\bar{\theta}_{n+1})f^{D=4}_n
\end{equation} 
where the advantage here is that $f^{D=4}_n$ can be directly lifted to the six dimensional $f^{D=6}_n$ defined in~\cite{Dennen:2010dh}. A side result of this analysis is the realization that the dual symmetry in six dimensions can be trivially extended from the original bosonic dual conformal to ``half" of the dual ${\it super}$ conformal symmetry. 

The amplitude with manifest dual conformal symmetry can be straight forwardly derived at four and five-point via half-Fourier transform of the fermionic variables. The half-Fourier transformed result can be written down at arbitrary $n$ for the MHV-$\overline{\rm MHV}$ amplitudes. From the five point amplitude we are able to obtain the basic $(\theta,\bar\theta)$ dependent dual conformal covariant function $\mathcal{R}_{r,st}$, which can serve as building blocks for higher point amplitudes. Equipped with this dual conformal covariant function, one can recursively reconstruct higher point amplitudes~\cite{Yutin} similar to Drummond and Henns result for the chiral representation~\cite{Drummond:2008cr}.

The appeal of the non-chiral representation is that it can be continued to six dimensions. The uplifting is continuous, since both $f^{D=4}_n$ and $f^{D=6}_n$ both have the same symmetry and are purely expressed in terms of momenta and supermomenta inner products, i.e. no spinor products. Obtaining the six-dimensional amplitude is then equivalent to four-dimensional massive ones, therefore a recursive solution for non-chiral amplitudes is equivalent to a recursive solution for massive amplitudes. The caveat is such a solution will be more complicated than a chiral formulation, since an $n$-point massless amplitude in the non-chiral representation is as complicated as  an $N^{n-4}MHV$ amplitude in the chiral representation. Indeed, one can compare the complexity of the closed form formula of the MHV amplitude given here with the all $n$-formula derived from the chiral approach in~\cite{Michael}. 

Finally, we note that this new representation is very similar to the dual symmetry in $\mathcal{N}=6$ super Chern-Simons theory~\cite{ABJM}, whose dual superspace is also non-chiral and requires additional bosonic coordinates that carries R-indices. It has been proven that the general $n$-point amplitude is OSp(6$|$4) DSC invariant at tree and at the level of (cut constructible parts) integrands~\cite{Lee:2010du,Gang:2010gy}.  It was conjectured that the presence of these three bosonic coordinates implies that the string sigma model is self-dual under fermionic T-duality~\cite{FermiT1,FermiT2}, if one T-dualizes three directions in both AdS$_4$ and CP$^3$. However, despite several attempts~\cite{Adam:2009kt, Grassi:2009yj,Adam:2010hh, Bakhmatov:2010fp,Dekel:2011qw}, such self duality was not established due to emergence of singularities. On the other hand, the appearance of the $y$ coordinates is well understood in four-dimensions. Indeed the non-chiral dual symmetry in four dimensions is understood as the self-duality of AdS$_5\times$S$_5$ background~\cite{FermiT1} under eight bosonic and fermionic T-dualities along the directions specified exactly by the projective superspace coordinates. A careful comparison might shed light on the nature of obstruction in three dimensions. An immediate interesting question would be the stringy origin of the fact that the amplitudes, or Wilson loop, are non-vanishing only on the support of  
$$\delta^4(y_{i,i+1}-\frac{\bar\theta_{ii+1}x_{i+1i+2}\theta_{ii+1}}{x^2_{ii+2}})\,.$$
While this constraint is straightforward from the on-shell amplitude point of view, its geometric origin in the sigma model is unclear. Further understanding of this non-chiral fermionic T-duality might also shed light on the correlator Wilson-loop duality~\cite{Alday:2010zy}, where these $y$ coordinates appear in the correlation functions as well.

\section{Acknowledgements}
The author would like to thank Harald Ita, Zvi Bern, David Skinner, Lionel Mason, Simon Caron-Huot, Cristian Vergu, Arthur Lipstein and Warren Siegel for useful discussions. This work is supported by the US Department of Energy under contract DE-FG03-91ER40662 and in part by the National Science Foundation under Grant No. NSF PHY05-51164..
\appendix

\end{document}